\documentclass[smallextended]{svjour3}
\smartqed  % flush right qed marks, e.g. at end of proof
\usepackage{amsfonts}
\usepackage{graphicx}
\usepackage{amssymb,amsxtra,amsmath}
\usepackage{latexsym}
\usepackage{booktabs}
\usepackage{cite}
\usepackage{url}
\usepackage{color}
\usepackage{tikz}
\usepackage{fancyvrb}
\usepackage{graphicx}
\usepackage{caption}
\usepackage{subcaption}
\usepackage[section]{placeins}

\newcommand{\vX}{ \mathbf{x} }
\newcommand{\vY}{ \mathbf{y} }

\begin{document}

\title{A Radial Basis Function (RBF) Method for the Fully Nonlinear 1D Serre Green--Naghdi Equations}
\titlerunning{RBF Method For the Fully Nonlinear 1D Serre Green--Naghdi Equations}

\author{Maurice S. Fabien}
\authorrunning{Maurice S. Fabien}

\institute{Maurice S. Fabien \at
              Department of Applied Mathematics, Rice University, Houston, TX 77005 \\
              \email{fabien@rice.edu}           %  \\
}
\date{June 2014}
\maketitle
\begin{abstract}
In this paper, we present a method based on Radial Basis Functions (RBFs) for numerically solving the fully nonlinear 1D Serre Green-Naghdi equations.  The approximation uses an RBF discretization in space and finite differences in time; the full discretization is obtained by the method of lines technique.  For select test cases the approximation achieves spectral (exponential) accuracy.  Complete \textsc{matlab} code of the numerical implementation is included in this paper (the logic is easy to follow, and the code is under 100 lines).

\end{abstract}

\section{Introduction}
\label{sec:intro}

\noindent  A spectral method that has gained attention recently is based on the so--called \textit{radial basis functions} (RBFs).  RBFs are relatively new, first being studied by Roland L. Hardy in the 1970s, and not gaining significant attention until the late 1980s.  Originally RBFs were used in scattered data modeling, however, they have seen a wide range of applications including differential equations.

RBFs offer a number of appealing properties, for instance they are capable of being spectrally accurate, they are a type of meshfree method, they are able to naturally produce multivariate approximations, and there is flexibility with the choice of a family of basis functions.  RBFs can avoid costly mesh generation, scale to high dimensions, and have a diverse selection of basis functions with varying smoothness.
\section{Introduction to RBF interpolation and differentiation}
\label{sec:rbf_review}

In this section we provide an introduction to RBF interpolation and differentiation.
\begin{definition}
Let $\Omega \subseteq \mathbb{R}^d$, and $\|\cdot\|_2$ is the  usual Euclidean norm on $\mathbb{R}^d$.  A function ${\bf\Phi}: \Omega \times \Omega \to \mathbb{R}$ is said to be \textbf{radial} if there exists a function $\varphi: [0,\infty) \to \mathbb{R}$, such that 
$$
\Phi(\vX,\vY) = \varphi(r),
$$
where $r= \lVert \vX - \vY \rVert_2$.
\label{def1}
\end{definition}
\begin{definition}
The function $\varphi(r)$ in definition~\ref{def1} is called a \textbf{radial basis function}.  That is, a radial basis function is a real--valued function whose value depends only on the distance from some point $\vec{y}$ called a \textit{center}, so that $\varphi(\vec{x},\vec{y}) =\varphi(\lVert \vec{x}-\vec{y} \rVert_2)$.  In some cases radial basis functions contain a free parameter $\epsilon$, called a \textit{shape parameter}.
\label{def2}
\end{definition}

Suppose that a set of scattered node data is given: 
$$
S= \{(\vX_i,f_i): \vX_i\in \Omega,\, f_i\in \mathbb{R},\text{ for } i=1,2,\ldots, N\},
$$
where $f:\Omega \to \mathbb{R}$ with $f(\vX_i)=f_i$.
Then, an RBF interpolant applied to the scattered node data takes the following form:
\begin{align}
g(\vX) = \sum_{i=1}^N \alpha_i \phi(\|\vX - \vX_i\|).
\label{eq:srbfinterp}
\end{align}
The unknown linear combination weights $\{\alpha_i\}_{i=1}^{N}$ can be determined by enforcing $g |_{S} = f|_{S}$. This results in a linear system:
\begin{align}
\underbrace{
\begin{bmatrix}
\phi(||\vX_1 - \vX_1||_2) & \phi(||\vX_1 - \vX_2||_2) & \hdots & \phi(||\vX_1 - \vX_N||_2)  \\
\phi(||\vX_2 - \vX_1||_2) & \phi(||\vX_2 - \vX_2||_2) & \hdots & \phi(||\vX_2 - \vX_N||_2) \\
\vdots & \vdots & \ddots & \vdots \\
\phi(||\vX_N - \vX_1||_2) & \phi(||\vX_N - \vX_2||_2) & \hdots & \phi(||\vX_N - \vX_N||_2) 
\end{bmatrix}}_{{\bf A}}
\underbrace{
\begin{bmatrix}
\alpha_1 \\
\alpha_2 \\
\vdots \\
\alpha_N 
\end{bmatrix}}_{ {\bf \alpha}}
=
\underbrace{
\begin{bmatrix}
f_1 \\
f_2 \\
\vdots \\
f_N
\end{bmatrix}}_{ {\bf f}}.
\label{eq:rbf_linsys}
\end{align}
The matrix $\bf A$ is sometimes called the \textit{RBF interpolation} or \textit{RBF system} matrix.  The RBF system matrix is always nonsingular for select RBFs $\phi$.  For instance the completely monotone multiquadratic RBF leads to an invertible RBF system matrix, and so do strictly positive definite RBFs like the inverse multiquadratics and Gaussians (see \cite{Micchelli} and \cite{mf} for more details).

There is flexibility in the choice of a RBF.  For instance, common RBF choices are: compactly supported and finitely smooth, global and finitely smooth, and global, infinitely differentiable (comes with a free parameter).  Table~\ref{table_local1} has a collection of some popular RBFs to illustrate the amount of variety there is in the selection of an RBF.  Optimal choices for RBFs are still a current area of research.  The wide applicability of RBFs makes the search for an optimal RBF challenging.  In order to keep the spectral accuracy of RBFs and the inveribility of system matrix in~(\ref{eq:rbf_linsys}), we will use the Gaussian RBFs (displayed in Table~\ref{table_local1}).

\begin{figure}[htb!]
\centering
\hspace*{-3ex}
\begin{subfigure}{.4\textwidth}
  \centering
  \centerline{\includegraphics[trim = 10mm 80mm 20mm 85mm, clip, scale = 0.45]{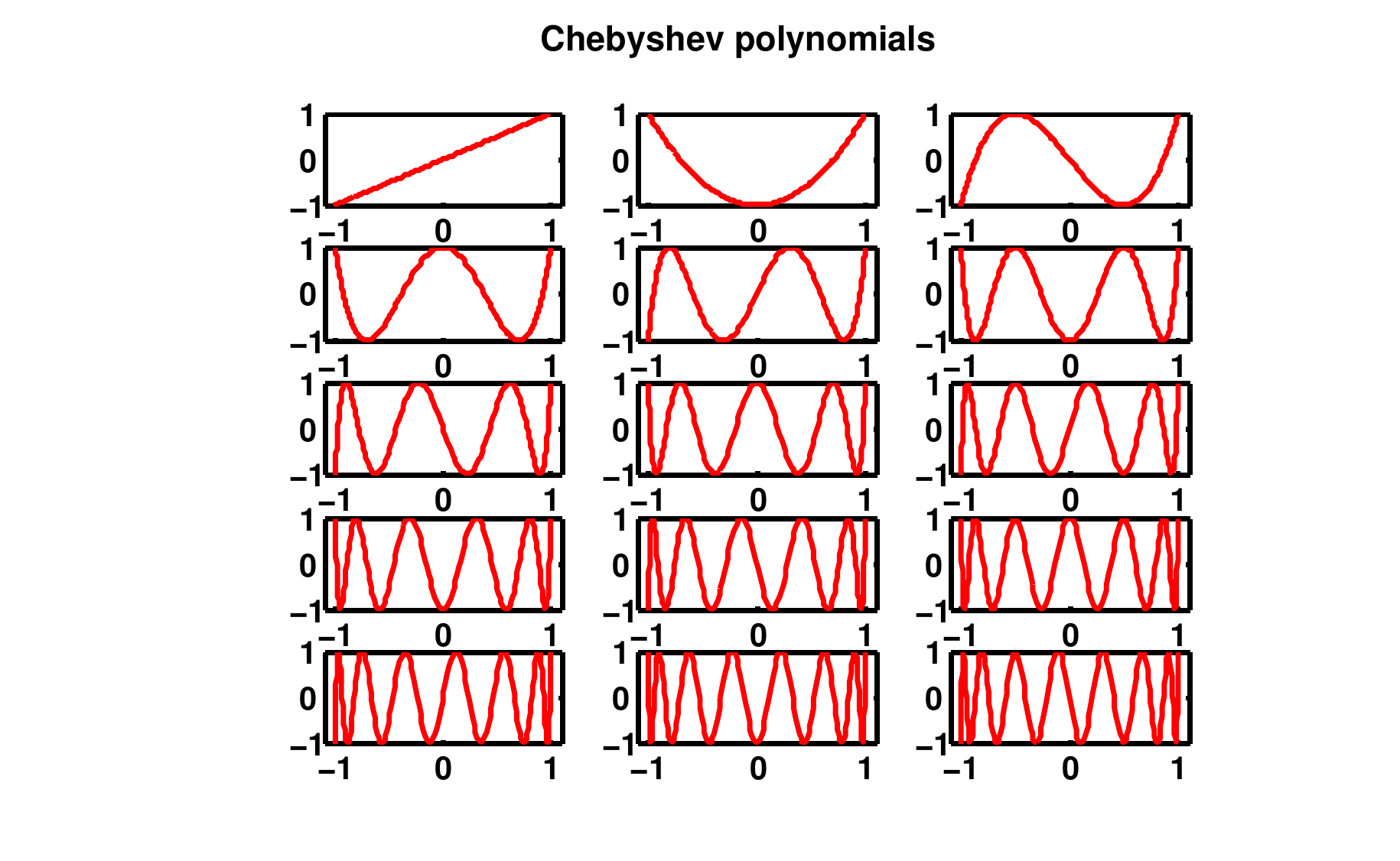}}
  \caption{ Chebyshev polynomials }
  \label{fig:rbf-general1}
\end{subfigure}%
\hspace*{10ex}
\begin{subfigure}{.4\textwidth}
  \centering
  \centerline{\includegraphics[trim = 10mm 80mm 20mm 85mm, clip, scale = 0.45]{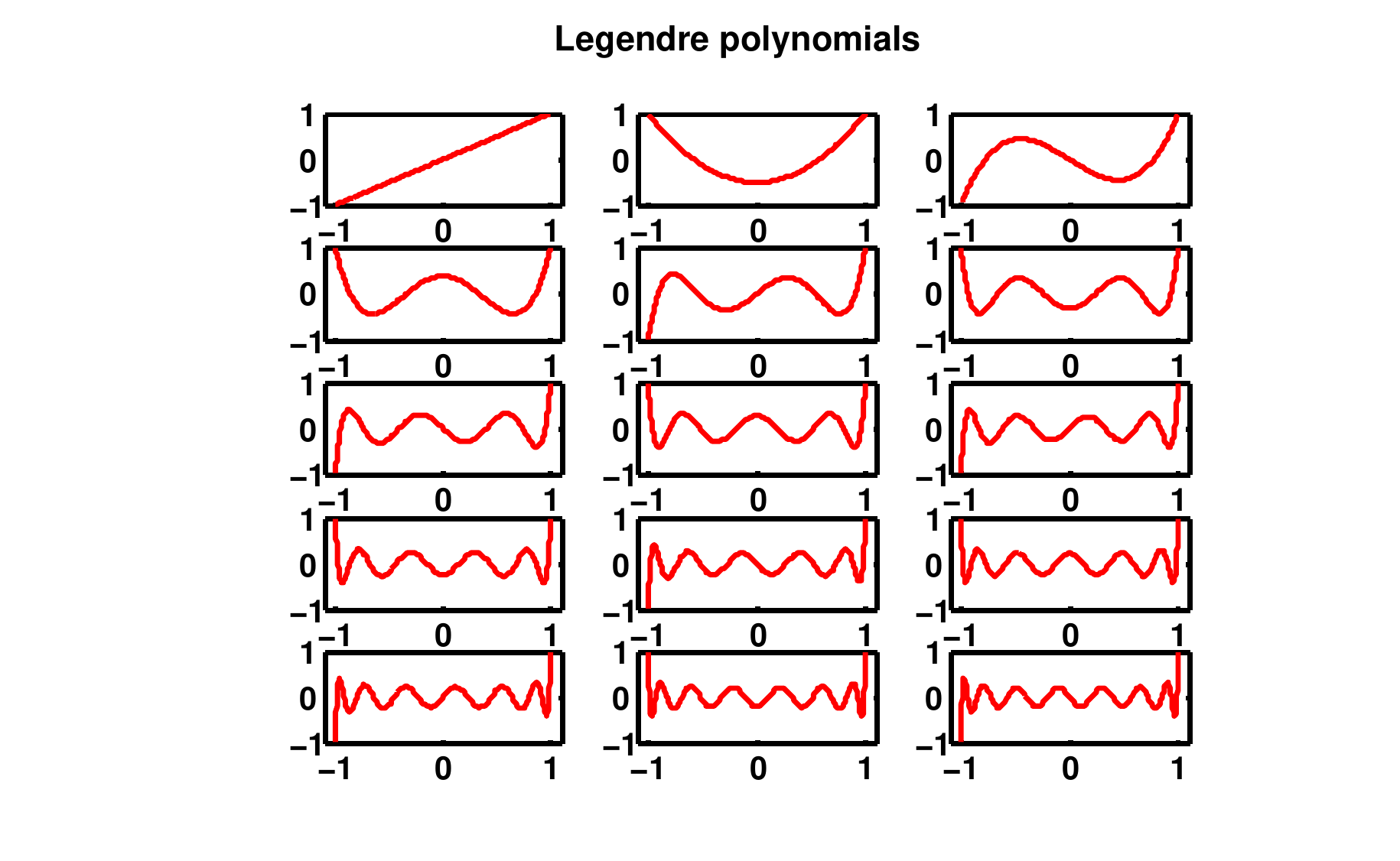}}
  \caption{ Legendre polynomials }
  \label{fig:rbf-general2}
\end{subfigure}
\hspace*{-3ex}
\begin{subfigure}{.4\textwidth}
  \centering
  \centerline{\includegraphics[trim = 10mm 80mm 20mm 85mm, clip, scale = 0.45]{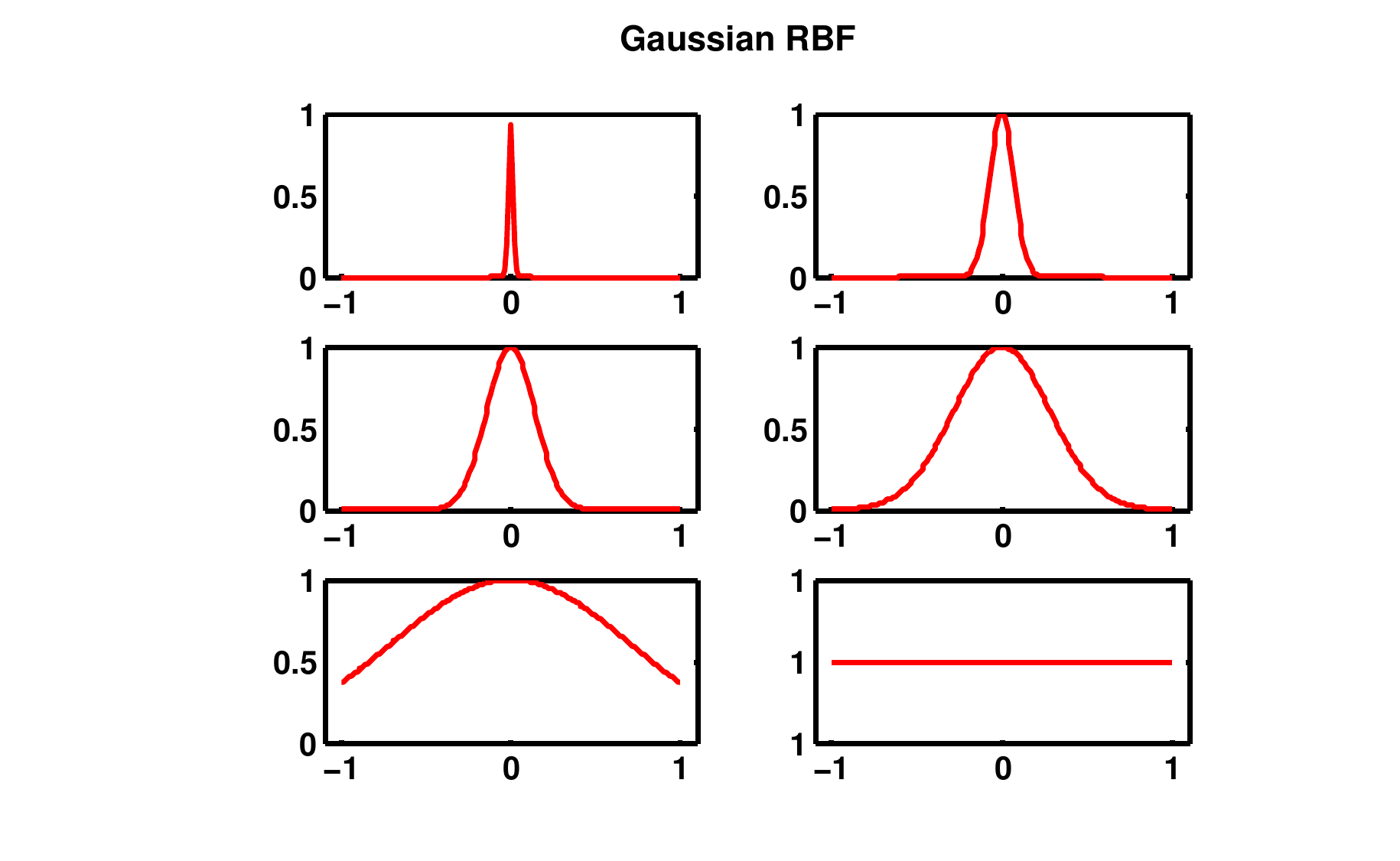}}
  \caption{ Gaussian RBF }
  \label{fig:rbf-general3}
\end{subfigure}%
\hspace*{10ex}
\begin{subfigure}{.4\textwidth}
  \centering
  \centerline{\includegraphics[trim = 10mm 80mm 20mm 85mm, clip, scale = 0.45]{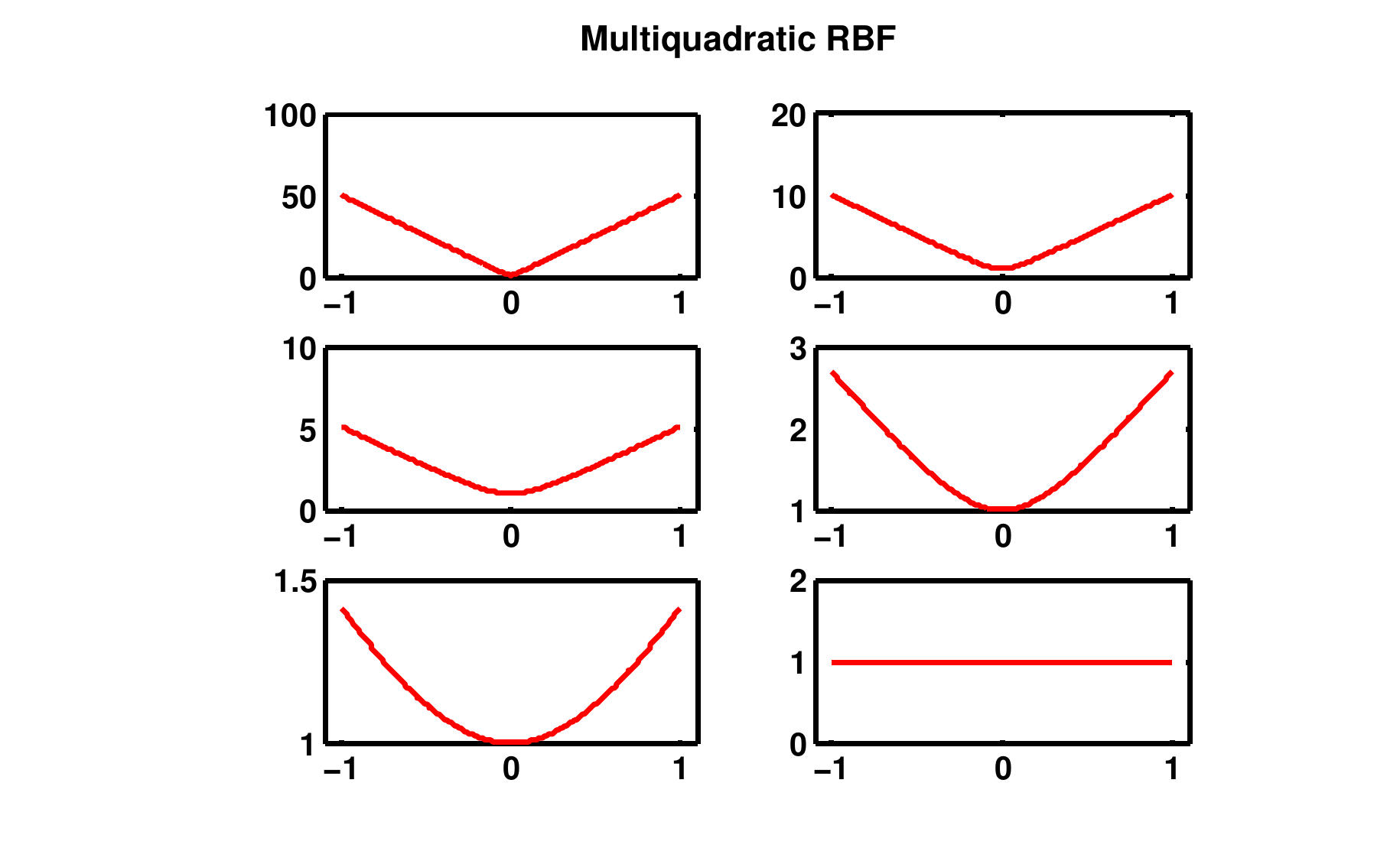}}
  \caption{ Multiquadratic RBF }
  \label{fig:rbf-general4}
\end{subfigure}
\caption{ Shape parameters used are from the set $\{50, 10, 5, 2.5, 1, 10^{-8}\}$.}
\label{fig:rbf-general}
\end{figure}

\begin{center}
\begin{table}[h]
\centering
\renewcommand*\arraystretch{1.4}
\begin{tabular}{|l|l|l|}
\hline 
Name of RBF             & Abbreviation  & Definition\\ 
\hline 
{\textcolor{red}{\textit{Smooth, global}}} &               & \\ 
\hline 
Multiquadratic          & MQ            & $\sqrt{1+(\epsilon r)^2}$\\ 
%\hline 
Inverse multiquadratic  & IMQ           & $(1+(\epsilon r)^2)^{-1/2}$\\ 
%\hline 
Inverse quadratic       & IQ           & $\frac{1}{1+(\epsilon r)^2}$\\ 
%\hline 
Gaussian                & GA            & $e^{-(\epsilon r)^2}$\\ 
\hline 
{\textcolor{blue}{\textit{Piecewise smooth, global}}}&      & \\ 
\hline 
Cubic                   & CU            & $|r|^3$\\ 
%\hline 
Quartic                   & QUA           & $|r|^4$\\ 
%\hline 
Quintic                   & QUI           & $|r|^5$\\ 
%\hline 
Thin plate spline type, order $k$ & TPS           & $|r|^{2k} \log{|r|}$\\ 
\hline 
{\textcolor{green}{\textit{Piecewise smooth, compact}}}&      & \\ \hline 
Wendland type, order 2&     W2           & $(1-\epsilon r)^4 (4 \epsilon r + 1)$ \\ 
Order 4&     W4           & $(1-\epsilon r)^6 ((35/3)(\epsilon r)^2 + 6\epsilon r +1)$ \\ 
Order 6&     W6           & $(1-\epsilon r)^8 (32(\epsilon r)^3 + 25(\epsilon r)^2 + 8\epsilon r + 1)$ \\ 
\hline 
\end{tabular}
\caption[Popular RBF functions (this table is adapted from \cite{Fornberg2}).]{Popular RBF functions.}%
\label{table_local1}
\end{table}
\end{center}
It should be noted that many of the RBFs contain a shape parameter $\epsilon$.  These parameters have a lot of control because each RBF center is assigned its own shape parameter (if so desired).  The shape parameter modifies the ``flatness'' of a RBF -- the smaller $\epsilon$ is, the flatter the RBF becomes.  Figures~\ref{fig:rbf-general3} and~\ref{fig:rbf-general4} visualizes this behavior.  In Figures~\ref{fig:rbf-general1} and~\ref{fig:rbf-general2} a contrasting behavior is shown for polynomial basis functions.  As the degree of the polynomial basis increases, the basis functions become more and more oscillatory.  

The shape parameter can play a substantial role in the accuracy of approximations.  A considerable amount of research has gone into the study of flat RBF interpolants.  Flat RBF interpolants (global, infinitely differentiable) are interesting because spectral accuracy is obtainable in the limit as $\epsilon \to 0$.  Interestingly enough, RBFs in the limit as $\epsilon \to 0$ have been shown to reproduce the classical pseudospectral methods (Chebyshev, Legendre, Jacobi, Fourier, etc.).  More precise statements and further details can be obtained in \cite{fornbergaccuracy}, \cite{Sarra}, \cite{fornberg2004some}, and \cite{driscoll2002interpolation}.

Even though RBF interpolants (global, infinitely differentiable) are capable of spectral accuracy in theory, and give rise to an invertible RBF system matrix, there are some computational considerations to be aware of.  First, there is no agreed upon consensus for selecting RBF centers and shape parameters.  Further, solving the linear system~(\ref{eq:rbf_linsys}) directly in practice (often called \textit{RBF--Direct}) is an unstable algorithm (see \cite{rbf_stable}).  This is due to the trade off (uncertainty) principle; as $\epsilon \to 0$, the system matrix becomes more ill conditioned.  Two stable algorithms currently exist for small shape parameters, Contour--Pad\'{e} and RBF--QR methods (see \cite{mf} and \cite{rbf_stable}).

\begin{figure}[htb!]
\centering
\hspace*{-3ex}
\begin{subfigure}{.4\textwidth}
  \centering
  \centerline{\includegraphics[trim = 10mm 80mm 20mm 85mm, clip, scale = 0.45]{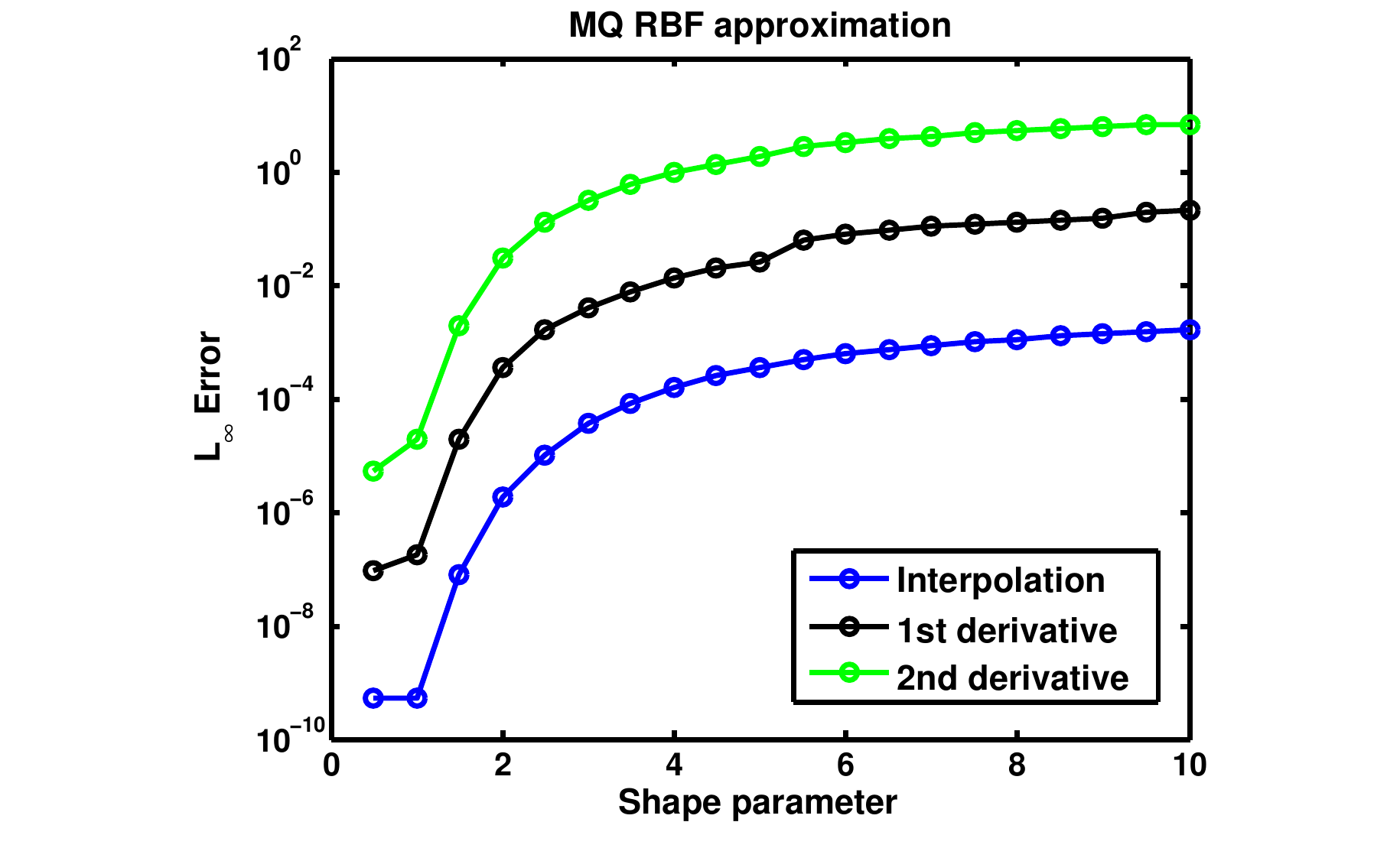}}
  \caption{ Approximation vs shape parameter }
  \label{fig:shp1}
\end{subfigure}%
\hspace*{10ex}
\begin{subfigure}{.4\textwidth}
  \centering
  \centerline{\includegraphics[trim = 10mm 80mm 20mm 85mm, clip, scale = 0.45]{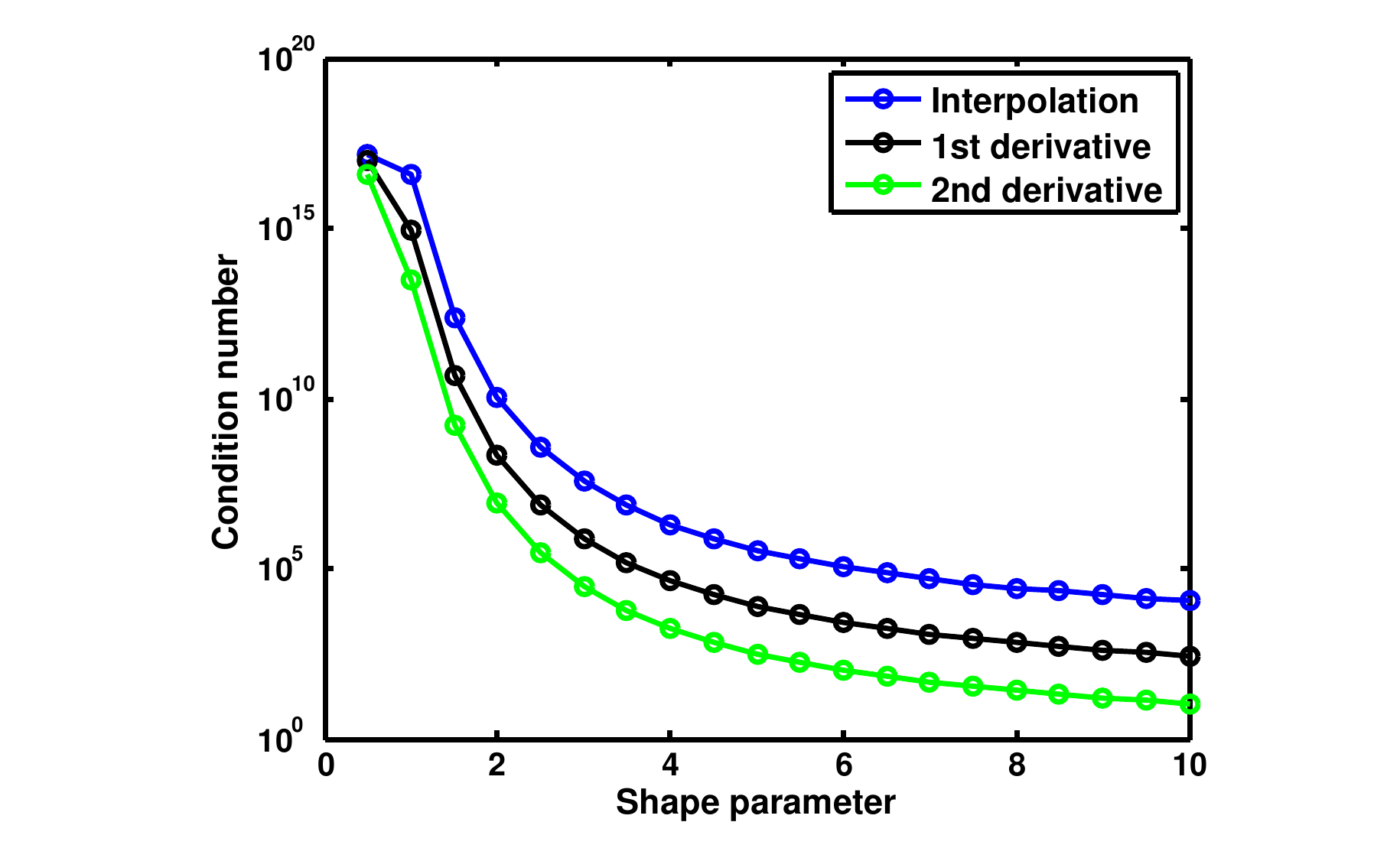}}
  \caption{ Shape parameter vs condition number }
  \label{fig:shp2}
\end{subfigure}
\caption{ Effect of shape parameter for RBF approximation.  Function used is $f(x)=\text{sech}^2(x)$ on $[-1,1]$ with $N=50$.  The interval $[-1,1]$ divided into 25 evenly spaced points provides the RBF centers for this example.  In Figure~\ref{fig:shp2} the evaluation matrices of first and second order are used (black and green lines respectively).}
\label{fig:shp}
\end{figure}

When RBF--Direct is used, selecting $\epsilon \ll 1$ is not always beneficial computationally.  This is due to the linear dependency between the shape parameter and the condition number of the system matrix: as $\epsilon \to 0$, $\kappa ({\bf A}) \to \mathcal{O}(\epsilon^{-M})$ for large $M>0$ (see Figure~\ref{fig:shp}).  Many strategies currently exist for selecting shape parameters (with RBF--Direct in mind), the monograph in \cite{Sarra} has a plentiful coverage.  A popular strategy is based on controlling the condition number of the system matrix such that it is within a certain range, say $10^9\le \kappa ({\bf A}) \le 10^{15}$.  If the condition number is not within the desired range, a different $\epsilon$ is selected and the condition number of the new system matrix is checked.

RBF differentiation matrices depend on the system matrix, thus they inherit poor conditioning.  To make this relationship more clear, we investigate how to construct RBF differentiation matrices.  To approximate a derivative, we simply differentiate both sides of equation~(\ref{eq:srbfinterp}) to obtain
\begin{equation}
\frac{\partial}{\partial \vec{x}_i}(g(\vec{x})) =  \sum_{j=1}^{N}{\alpha_j \frac{\partial}{\partial \vec{x}_i}\phi ( \lVert \vec{x} - \vec{x}_j\rVert_2)}=\bigg[\frac{\partial}{\partial \vec{x}_i}{\bf A}\bigg]\vec{\alpha},
\label{rbf-der}
\end{equation}
for $i=1,2,\ldots , N$ and $\vec{\alpha}=[\alpha_1,\ldots,\alpha_N]^T$.
We define the first order evaluation matrix $\bf D_1$ such that
$$
({\bf D_1})_{ij} =\frac{\partial}{\partial \vec{x}_i}\phi ( \lVert \vec{x} - \vec{x}_j\rVert_2),~~~i,j=1,2,\ldots N.
$$
If we select a RBF that gives rise to a nonsingular interpolation matrix, equation~(\ref{eq:rbf_linsys}) implies that $\vec{\alpha} = {\bf A}^{-1}\vec{f}$.  Then, from equation~(\ref{rbf-der}), we have
$$
\frac{\partial}{\partial \vec{x}_i}(g(\vec{x})) = \bigg[\frac{\partial}{\partial \vec{x}_i}{\bf A}\bigg]\vec{\alpha},
~~~
\text{ for }i=1,2,\ldots, N\implies
[{\bf D_1}]{\bf A}^{-1}\vec{f},
$$
and we define the \textit{differentation matrix} $\bf D_x$ to be
\begin{equation}
\bf D_x = {\bf D_1}{\bf A}^{-1}.
\label{rbf-der2}
\end{equation}
Following this process one can easily construct differentiation matrices of arbitrary order:
\begin{equation}
\frac{\partial^{(m)}}{\partial \vec{x}_i^{(m)}}(g(\vec{x})) = \bigg[\frac{\partial^{(m)}}{\partial \vec{x}^{(m)}_i}{\bf A}\bigg]\vec{\alpha}
=
\bigg[\frac{\partial^{(m)}}{\partial \vec{x}^{(m)}_i}{\bf A}\bigg]{\bf A}^{-1}\vec{f}.
\label{rbf-der3}
\end{equation}
Thus, the $m$th order RBF differentation matrix is given by
\begin{equation}
\bigg[{\bf D_m}\bigg]{\bf A}^{-1}.
\label{rbf-der4}
\end{equation}
Note that equation~(\ref{rbf-der2}) is a \textit{matrix system}, not a linear system.  The unknown variable ($\bf D_x$) in the equation ${\bf D_x}{\bf A}={\bf D_1}$ is a matrix.

In practice the matrix ${\bf A}^{-1}$ is never actually formed, a matrix system solver is used instead.  In \textsc{matlab} this can be done by the forward slash operator, or for singular or near singular problems the pseudoinverse.  The differentiation matrices derived above are called \textit{global}, since they use information from every center.

\subsection{Local RBF differentiation (RBF Finite Differences)}

This idea of local differentiation has been applied to RBFs -- and it is very popular in the RBF literature, especially when concerning time dependent PDEs.  Local RBFs, also known as RBF--FD (radial basis function finite differences) have produced a lot of interest due to their interpolation and differentiation matrix structure.  The interpolation and differentiation matrices generated by local RBFs have a controllable amount of sparsity.  This sparsity can allow for much larger problems and make use of parallelism.  

The main drawback of local RBFs is that spectral accuracy is no longer obtained.  In fact, the accuracy for local RBFs is dictated by the stencil used (the situation is similar for finite differences).  The literature for RBFs is currently leaning towards local RBFs, since global RBFs produce dense, ill conditioned matrices.  This drastically limits the scalability of global RBFs.  In this section we will examine the simplest of local RBFs, however, more advanced local RBFs can be found in \cite{Fornberg}.

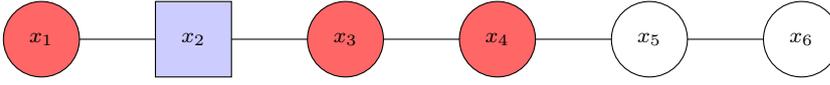
\begin{figure}[htb!]
\centering
\begin{tikzpicture}[node distance = \len, auto]
\tikzset{
    line/.style = {draw},
    block/.style = {rectangle, draw, text centered, minimum height=1em},
}
\draw (0,0) -- (9.5,0);
%\node at (-1,0)[draw,rectangle split, rectangle split horizontal,rectangle split parts=3,minimum height=1cm] {\nodepart{two}\shortstack{Predefined\\Process}};
\node at (-.5,0)[draw,circle,minimum height=1cm,fill=red!60] {$x_1$};
\node at (1.5,0)[draw,rectangle ,minimum width=1cm,minimum height=1cm,fill=blue!20] {$x_2$};
\node at (3.5,0)[draw,circle,minimum height=1cm,fill=red!60] {$x_3$};
\node at (5.5,0)[draw,circle,minimum height=1cm,fill=red!60] {$x_4$};
\node at (7.5,0)[draw,circle,minimum height=1cm,fill=white!10] {$x_5$};
\node at (9.5,0)[draw,circle,minimum height=1cm,fill=white!10] {$x_6$};
%\node at (11.5,0)[draw,circle,minimum height=1cm,fill=white!10] {$x_7$};
%\node at (13.5,0)[draw,circle,minimum height=1cm,fill=white!10] {$x_8$};
\end{tikzpicture}
\caption{Stencil example: a domain with 8 grid points and a stencil size of $m=4$.  At the node $x_2$ (blue square), the $m-1=3$ nearest neighbors (red circles) are used in calculations.}
\label{local_node}
\end{figure}

A RBF--FD stencil of size $m$ requires the $m-1$ nearest neighbors (see Figure~\ref{local_node}).  The local RBF interpolant takes the form
$$
I_m g(\vec{x}) = \sum_{k \in {\bf\mathcal{I}}_i}{\vec{\alpha}_k \phi (\lVert \vec{x} -\vec{x}_k\rVert_2)},
$$
where $\vec{x}$ contains the $N$ RBF centers, ${\bf\mathcal{I}}_i$ is a set associated with RBF center $i$ and whose elements are RBF center $i$'s $m-1$ nearest neighbors.  The vector $\vec{\alpha}$ is the unknown weights for the linear combination of RBFs.

These weights can be calculated by enforcing
$$
I_m g(\vec{x}_k)  =  g(\vec{x}_k), \text{ for each } k \in {\bf\mathcal{I}}_i.
$$
This results in $N$ linear systems of size $m\times m$, ${\bf B}\vec{\alpha}=\vec{g}$.  The vector $\vec{g}$ is given by $\vec{g}=[g(\vec{x}_1),g(\vec{x}_2),\ldots g(\vec{x}_N)]^T$.  The entries of ${\bf B}$ have a familiar form
$$
{\bf B}_{jk} = \phi (\lVert \vec{x}_j -\vec{x}_k\rVert_2),~~~j,k \in {\bf\mathcal{I}}_i.
$$
By selecting global infinitely differentiable RBFs (GA, MQ, IMQ, etc.), the matrix ${\bf B}$ is guaranteed to be nonsingular.  This implies that the coefficients on each stencil are uniquely defined.  Local RBF derivatives are formed by evaluating
$$
\mathcal{L} g(\vec{x}) = \sum_{j=1}^{N}{\vec{\alpha}_j \mathcal{L}\phi (\lVert \vec{x} -\vec{x}_j\rVert_2)},
$$
at a RBF center where the stencil is based, and $\mathcal{L}$ is a linear operator.  This equation can be simplified to
$$
\mathcal{L} g(\vec{x}_i) = \vec{h}^T \vec{\alpha},
$$
where $\vec{h}$ (a $m \times 1$ vector and $\vec{\alpha}$ contains the contains the RBF linear combination weights for the centers in $\mathcal{I}_i$) has components of the form
$$
(\vec{h})_i = \phi (\lVert \vec{x} -\vec{x}_i\rVert_2),~~~i=1,2,\ldots , m.
$$
It then follows that
$$
\mathcal{L} g(\vec{x}_i) = (\vec{h}^T{\bf B}^{-1})g \big|_{{\bf\mathcal{I}}_i}
=
(\vec{w}_i)g \big|_{{\bf\mathcal{I}}_i},
$$
and $\vec{w}_i=\vec{h}^T{\bf B}^{-1}$ are the stencil weights at RBF center $i$ ($\vec{w}_i$ multiplied by the function values provides the derivative approximation).  The $i$th row of the $m$th order local RBF differentiation matrix is then given by $({\bf W_m})_i = [\vec{w}_i]$.

\begin{figure}[htb!]
\centering
\hspace*{-3ex}
\begin{subfigure}{.4\textwidth}
  \centering
  \centerline{\includegraphics[trim = 10mm 80mm 20mm 85mm, clip, scale = 0.45]{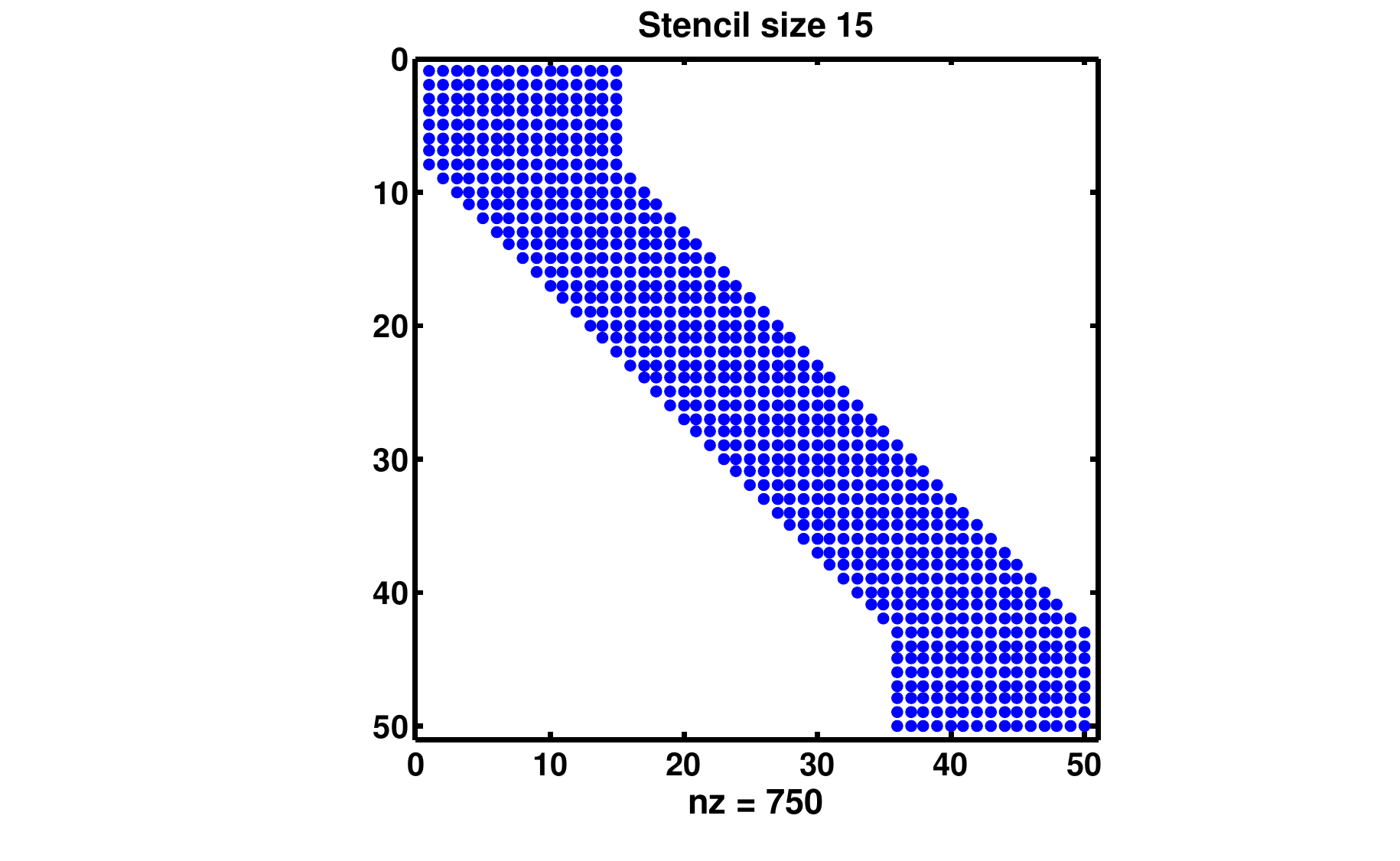}}
  \caption{ Stencil size 15 ($50 \times 50$) matrix }
  \label{fig:loc1}
\end{subfigure}%
\hspace*{10ex}
\begin{subfigure}{.4\textwidth}
  \centering
  \centerline{\includegraphics[trim = 10mm 80mm 20mm 85mm, clip, scale = 0.45]{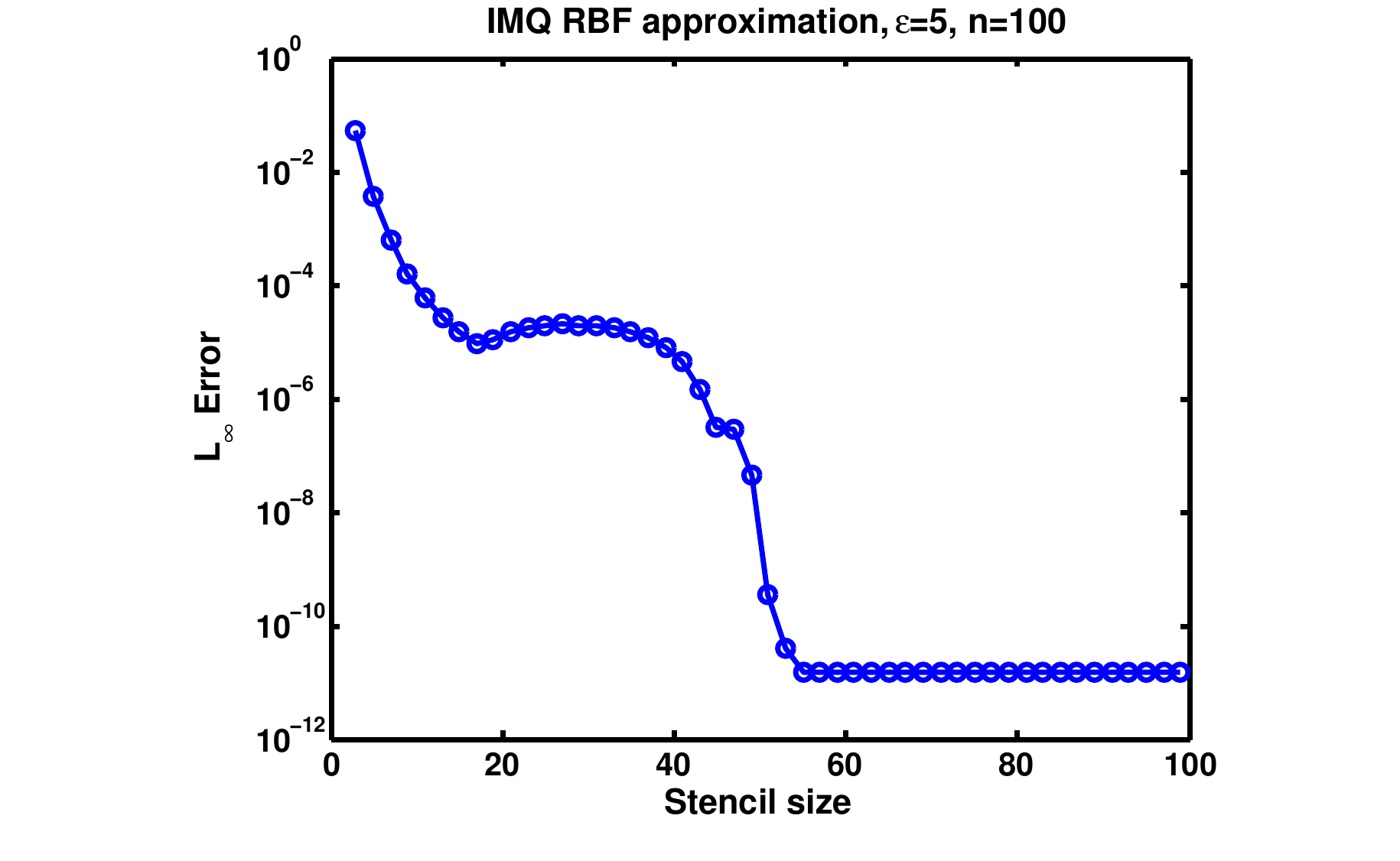}}
  \caption{ Effect of stencil size }
  \label{fig:loc2}
\end{subfigure}
\caption{ Sparsity patterns and the relationship between accuracy and stencil size.}
\label{fig:loc}
\end{figure}

Figure~\ref{fig:loc} has an illustration of sparsity patterns and the relationship between accuracy and stencil size.  The test problem is the the Runge function $1/(1 + 25x^2)$ on the interval $[-1, 1]$ (100 equally spaced points, and a blanket shape parameter of 5).

For higher dimensional problems RBF--FDs have been shown to be a viable alternative.  For examples see \cite{Fornberg}, \cite{bollig2012solution}, and \cite{Flyer}.  In this article, we only consider the one--dimensional Serre Green-Naghdi equations.  And for this particular application of one--dimensional RBFs, local differentation provides comparable results to the global case.

\section{Numerical Time Stepping Stability}
\label{sec:stability}
The spatial dimensions are to be discretized by RBFs.  To achieve a complete discretization the well known method of lines will be employed.  Details of this technique can be found in \cite{schiesser1991numerical}.  A rule of thumb for stability of the method of lines is to have the eigenvalues of the linearized spatial discretization operator scaled by $\Delta t$ to be contained in the stability region of the ODE solver invoked (see \cite{reddy1992stability}).  The Serre Green-Naghdi equations are nonlinear, so it is more convenient for spectral methods if an explicit ODE solver is used.  In this situation we would like the scaled eigenvalues to lie in the left half of the complex plane.
\vspace*{-5ex}
\begin{figure}[htb!]
\centering
\hspace*{-3ex}
\begin{subfigure}{.4\textwidth}
  \centering
  \centerline{\includegraphics[trim = 10mm 80mm 20mm 85mm, clip, scale = 0.45]{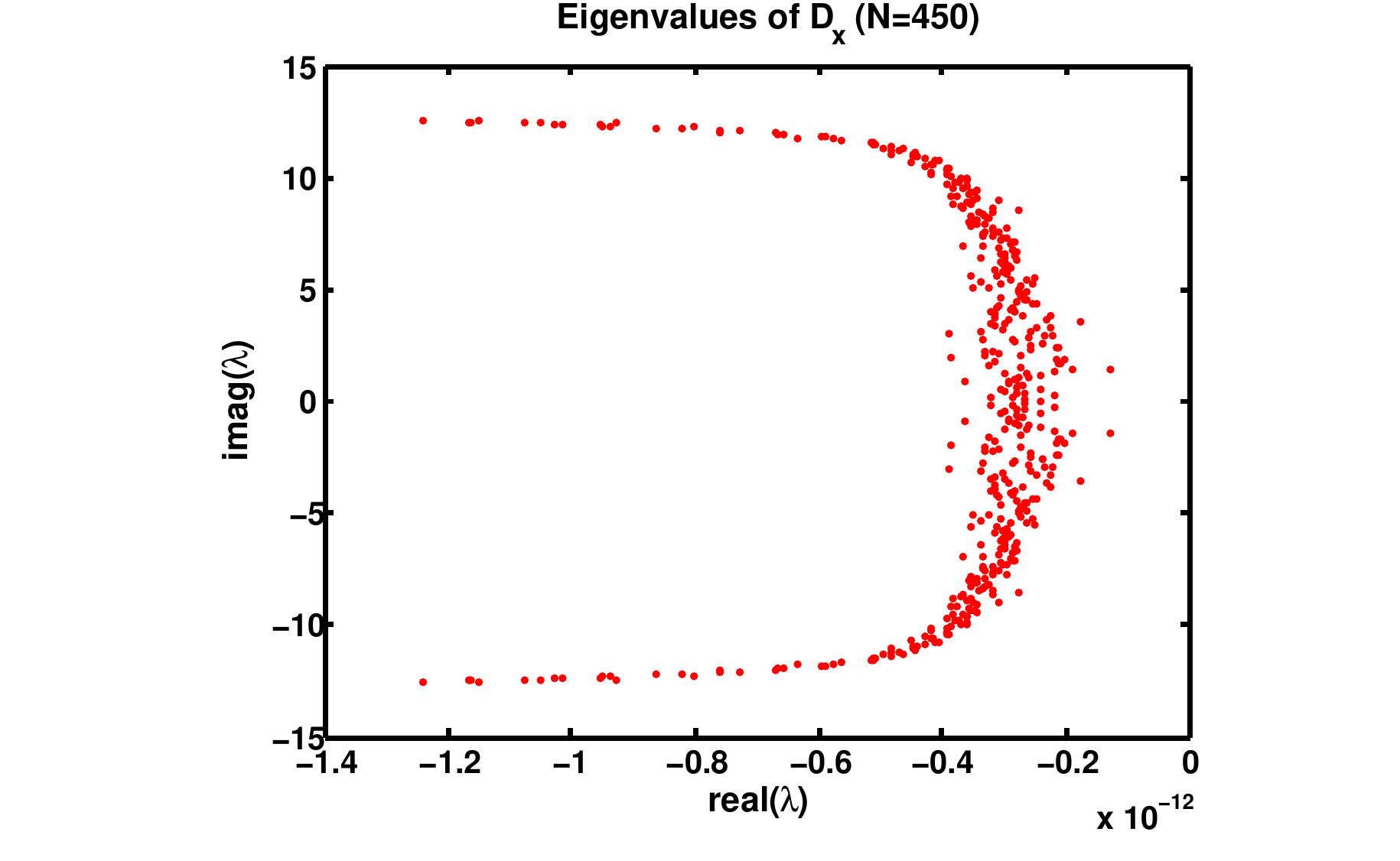}}
  \caption{ Eigenvalues of $D_x$ with $N=450$ }
  \label{fig:eig1}
\end{subfigure}%
\hspace*{10ex}
\begin{subfigure}{.4\textwidth}
  \centering
  \centerline{\includegraphics[trim = 10mm 80mm 20mm 85mm, clip, scale = 0.45]{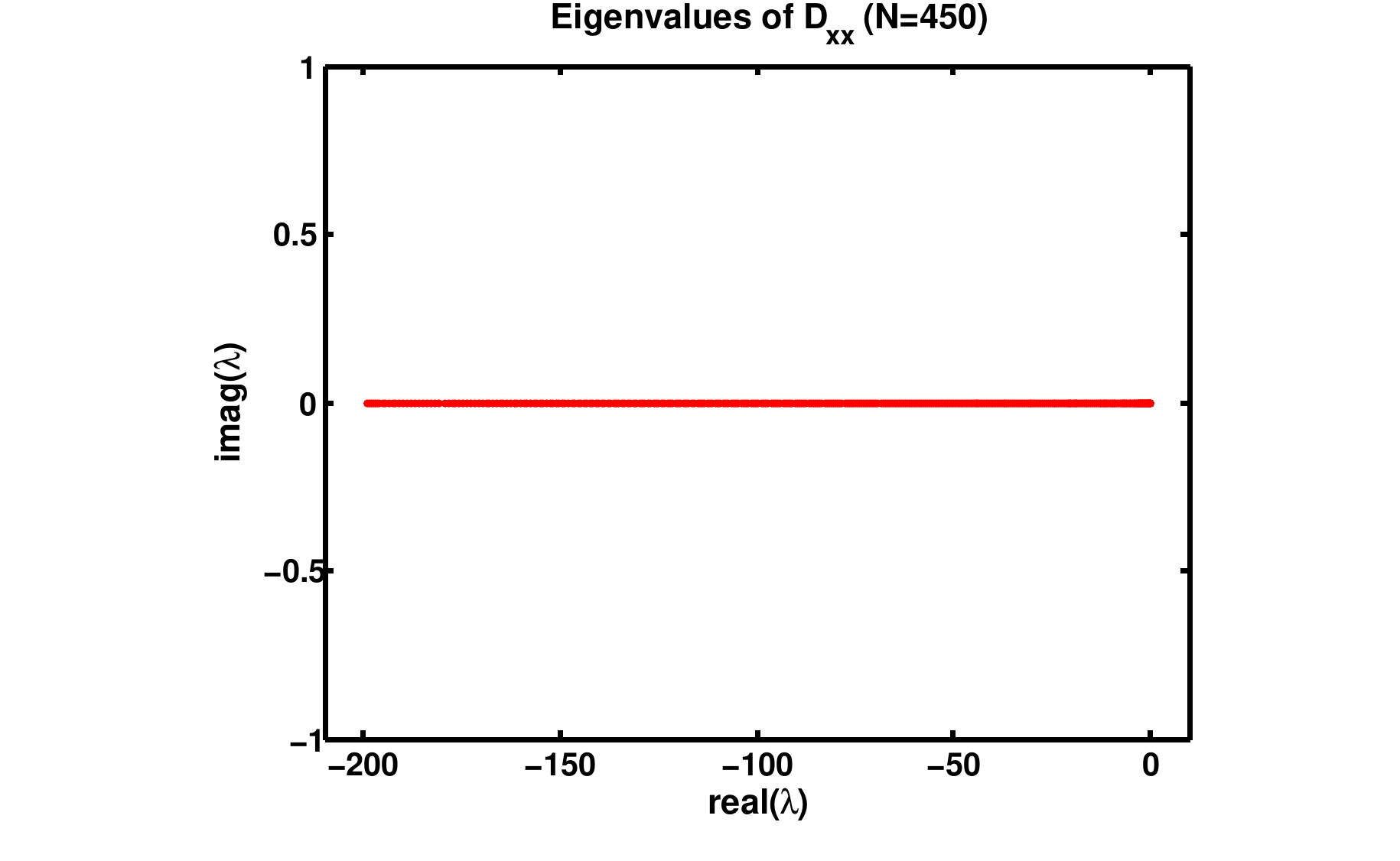}}
  \caption{ Eigenvalues of $D_{xx}$ with $N=450$ }
  \label{fig:eig2}
\end{subfigure}
\caption{ Eigenvalues of spatial differentation matrices for $N=450$ and $\epsilon=2$.  In figure~\ref{fig:eig2} the eigenvalues are clustered along the real axis.  Figures~\ref{fig:eig1} and~\ref{fig:eig2} are not scaled by $\Delta t$. }
\label{fig:eig}
\end{figure}

Achieving stability in the method of lines discretization is still being researched.  The eigenvalue location of RBF differentation matrices is irregular.  Altering $\epsilon$, or $N$ can cause the locations to adjust nontrivially.  For hyperbolic PDEs an answer to this difficultly has been found in the concept of \textit{hyperviscosity}.  Adding artificial viscosity has been shown in many cases to stabilize the numerical time stepping for hyperbolic PDEs.  For instance, in \cite{bollig2012solution} hypervisocity is used for the two--dimensional shallow water equations (with a RBF spatial discretization).  Also, in \cite{fornberg2011stabilization} hyperviscosity is applied to convective PDEs in an RBF setting.
\section{Discretization}
\label{sec:discretization}
In this section we discretize the fully nonlinear 1D Serre Green-Naghdi (SGN) equations.  This nonlinear hyperbolic PDE system is given by
\begin{align}
h_t + (uh)_x &=0 \label{eq_sgn0}\\
u_t + (0.5u^2 +gh)_x &=\beta h^{-1}[h^3(u_{xt}+uu_{xx}-u_x^2)]_x,
\label{eq_sgn1}
\end{align}
where $g$ is the acceleration due to gravity, and $\beta = 1/3$.  For spectral methods it is easier\footnote{Equation~(\ref{eq_sgn1}) creates difficulties for RBF spectral methods due to the term $u_{xt}$ located on the right hand side.} to work with the following equivalent system (see \cite{Dutykh2})
\begin{align}
\eta_t + [u(d+\eta)]_x &=0 \label{eq_sgn2-1}\\
q_t + [qu-0.5u^2 +g\eta -0.5(d+\eta)^2u_x^2]_x &=0 , \label{eq_sgn2-2}\\
q-u+\beta (d+\eta)^2u_{xx} + (d+\eta)\eta_xu_x &=0,  \label{eq_sgn2-3}
\end{align}
where $\eta=\eta(x,t)$ is the free surface elevation ($h(x,t)=d+\eta(x,t)$, we will assume $d$ is constant), $u=u(x,t)$ is the depth-averaged velocity, and $q=q(x,t)$ is a conserved quantity of the form $q(x,t)=uh-\beta [h^3u_x]_x$.  Equations~(\ref{eq_sgn2-1}) and~(\ref{eq_sgn2-2}) have time dependent derivatives, however, equation~(\ref{eq_sgn2-3}) has no time dependent derivatives.  Hence, the numerical strategy will be to evolve equations~(\ref{eq_sgn2-1}) and~(\ref{eq_sgn2-2}) in time, and at each time step the elliptic PDE~(\ref{eq_sgn2-3}) will be approximated.

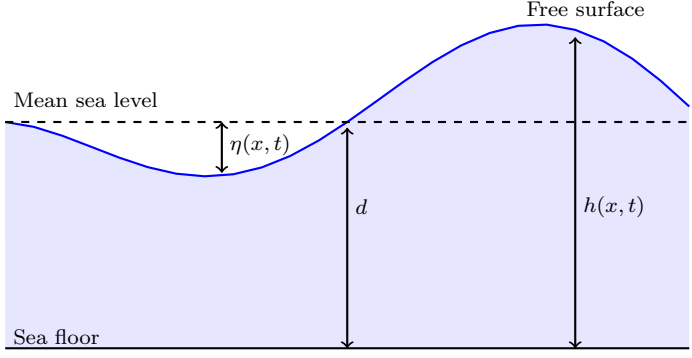
\begin{figure}[htb!]
\centering
\hspace*{-10ex}
\begin{tikzpicture}[scale=1.5]
	\fill[fill=blue!10] (-1,1) -- plot [domain=-3:3] (\x, {2+(sin(\x r)* ln(\x+4))/2}) -- cycle;
	\fill[fill=blue!10] (-1,1) -- (-3,2) parabola bend (-1,0.5) (3,0) -- cycle;
	\fill[fill=blue!10] (3,0) -- (-1,1) -- (-1,1) -- (3,1) -- cycle;
	\fill[fill=blue!10] (3,2.15) -- (0,2) -- (-1,1) -- (3,1) -- cycle;
	\fill[fill=blue!10] (-1,1) -- (-3,1) -- (-1,0) -- (3,0) -- cycle;
	\fill[fill=blue!10] (-1,1) -- (-3,0) -- (-1,0) -- (3,0) -- cycle;
	\fill[fill=blue!10] (-3,1) -- (-3,2) -- (-1,0.5) -- (3,0)-- cycle;
	\fill[fill=blue!10] (-3,0) -- (-3,2) -- (-1,0.5) -- (3,0)-- cycle;

    \draw[blue,thick,domain=-3:3] plot (\x, {2+(sin(\x r)* ln(\x+4))/2});
    \draw[black,thick,dashed]  (-3,2) -- (3,2);
    
    \draw[black,thick,-]  (-3,0) -- (3,0);

    \draw[black,thick,<->]  (0,0) -- (0,1.95);
    \draw (0,1.25) node[right] {$d$};
    
    \draw[black,thick,<->]  (2,0) -- (2,2.75);
    \draw (2,1.25) node[right] {$h(x,t)$};

    \draw[black,thick,<->]  (-1.1,2.0) -- (-1.1,1.55);
    \draw (-1.1,1.8) node[right] {$\eta(x,t)$};
    
    \draw (1.5,3.0) node[right] {Free surface};
    
    \draw (-3,0.1) node[right] {Sea floor};
    
    \draw (-3,2.2) node[right] {Mean sea level};
\end{tikzpicture}
\caption{Relevant variables for the SGN equations.}
\label{fig:ch_5_1_000}
\end{figure}

  The 1D SGN equations admit exact solitary wave solutions:
\begin{align}
\eta(x,t) & = a \cdot \text{sech}{(0.5\kappa (x-ct))^2}, \label{eq_sgn3-1}\\
u(x,t)    & = \frac{c\eta}{d+ \eta}, \label{eq_sgn3-2}
\end{align}
where $c= \sqrt{g(d+a)}$ is the wave speed, $a$ is the wave amplitude, and $(\kappa d)^2 = a / (\beta (d+a))$.  See Figure~\ref{fig:ch_5_1_000} for a visualization of relevant parameters.

\subsection{Global RBF spectral method}

We take a radial basis function approach.  To begin the collocation, partition the spatial domain (an interval in the 1D SGN case) as $x_1,x_2,\ldots, x_N$, and suppose that $N$ centers $y_1,y_2,\ldots y_N$ have been selected (for simplicity we take the centers to agree with the spatial domain partition).  Then the RBF interpolation and differentiation matrices need to be constructed.  The RBF interpolation matrix $\bf A$ has entries
$$
{\bf A}_{ij} = \phi (\lVert x_i - y_j \rVert_2).
$$
The first and second RBF order evaluation matrices are given by
$$
{\bf D_1}_{ij} = \frac{\partial} {\partial x_i} \phi (\lVert x_i - y_j \rVert_2),~~~
{\bf D_2}_{ij} = \frac{\partial^2} {\partial x_i^2} \phi (\lVert x_i - y_j \rVert_2),
$$
for $i,j = 1,\ldots , N.$  Then the first and second RBF differentiation matrices denoted by $\bf D_x$ and $\bf D_{xx}$, respectively, are defined as ${\bf D_x}  = {\bf D_1} {\bf A}^{-1}$ and ${\bf D_{xx}}  = {\bf D_2} {\bf A}^{-1}$.

\noindent  Let the variables $\vec x,$ ${\vec \eta},$ ${\vec q}$ and $,{\vec u} $ be given by
$$
{\vec x} = 
\begin{bmatrix}
x_1 \\
x_2 \\
\vdots \\
x_N
\end{bmatrix},
~~~
{\vec \eta} = 
\begin{bmatrix}
\eta(x_1,t) \\
\eta(x_2,t) \\
\vdots \\
\eta(x_N,t)
\end{bmatrix},
~~~
{\vec q} = 
\begin{bmatrix}
q(x_1,t) \\
q(x_2,t) \\
\vdots \\
q(x_N,t)
\end{bmatrix},
~~~\text{and}~~~
{\vec u} = 
\begin{bmatrix}
u(x_1,t) \\
u(x_2,t) \\
\vdots \\
u(x_N,t)
\end{bmatrix}.
$$
In addition, let $\vec{d} \in \mathbb{R}^N$ be a vector with $d$ in each component.  The variable $d$ represents the depth above the.  Given $\vec{p} \in \mathbb{R}^N$, let the function $\text{diag}: \mathbb{R}^N \to \mathbb{R}^{N\times N} $ be defined as
$$
(\text{diag}\{\vec{p}\})_{ij}=
\begin{cases}
\vec{p}_i,  \text{ if } i =j\\
0        , \text{ otherwise},
\end{cases}
~~~
\text{for   }
~~~
 i,j=1,2,\ldots ,N.
$$
  The equations~(\ref{eq_sgn2-1}), (\ref{eq_sgn2-2})~, and~(\ref{eq_sgn2-3}) can be expressed as a semidiscrete system
\begin{align}
 \vec{\eta}_t + [{\bf D_x}(\vec{d} + \vec{\eta})] &=\vec{0} \label{eq_sgn4-1}\\
\vec{q}_{t} + {\bf D_x}[ \vec{q} \otimes \vec{u}-0.5(\vec{u})^2 +g \vec{\eta} -0.5 \text{diag}\{ ( \vec{d} +\vec{\eta})^2\} ({\bf D_x} \vec{u})^2] &=\vec{0},\label{eq_sgn4-2}\\
\vec{q} - \vec{u}+ \text{diag}\{\beta (\vec{d} +\vec{\eta})^2 \}({\bf D_{xx}}\vec{u})+ \text{diag}\{(\vec{d} + \vec{\eta}) \otimes ({\bf D_x}\vec{\eta})\}({\bf D_x}\vec{u}) &=\vec{0}.\label{eq_sgn4-3}
\end{align}
The $\otimes$ operator is element--wise multiplication.  Also, in equations~(\ref{eq_sgn4-2}) and~(\ref{eq_sgn4-3}) the square operator on vectors is element--wise.  The method of lines can be employed from here to fully discretize the 1D Serre-Green Nagdhi equations.  Equations~(\ref{eq_sgn4-1}) and~(\ref{eq_sgn4-2}) are of evolution type, and equation~(\ref{eq_sgn4-3}) can be treated like an elliptic PDE in the variable $\vec{u}$.  To initialize $\eta$ and $u$, equations~(\ref{eq_sgn3-1}) and~(\ref{eq_sgn3-2}) are used.  The differentiation matrix ${\mathbf{D_x}}$ is used to initialize $q(x,t) = uh-\beta [h^3u_x]_x$.  

	The sample code given in {{Appendix A}} uses \textsc{matlab}'s {\tt{ode113}} (variable order Adams--Bashforth--Moulton PECE solver).  We take full advantage of the error tolerance specifications \textsc{matlab}'s ODE solvers provide.  We do this to attempt to match the accuracy of the temporal discretization with the spatial discretization.

Below is a high level algorithm for the implementation of the RBF spectral method for 1D SGN.

\begin{center}
\tikzstyle{mybox} = [draw=red, fill=blue!15, very thick,
    rectangle,  inner sep=10pt, inner ysep=20pt]
\tikzstyle{fancytitle} =[fill=red!20, text=black]
\begin{tikzpicture}
\node [mybox] (box){%
    \begin{minipage}{1\textwidth}
		\begin{itemize}
			\item  Step 1:  Select a RBF, RBF centers, collocation points, and a time step.
			\item  Step 2:  Construct the required RBF interpolation and differentation matrices outside of the main time stepping loop.
			\item  Step 3:  Use an ODE solver to advance the coupled evolution equations~(\ref{eq_sgn4-1}) and~(\ref{eq_sgn4-2}) to time level $t_{n+1}$.
			\item  Step 4:  Solve the linear system in equation~(\ref{eq_sgn4-3}) at the time level $t_n$.  This updates the variable $u$ to time level $t_{n+1}$.
			\item  Step 5:  Repeat steps 3 and 4 until final time is reached.		
		\end{itemize}
    \end{minipage}
};
\node[fancytitle, right=10pt] at (box.north west) {Algorithm (RBF spectral method) for 1D SGN};
%\node[fancytitle, rounded corners] at (box.east) {$\clubsuit$};
\end{tikzpicture}%
\end{center}
\newpage
\subsection{Test Cases for the 1D Serre--Green Nagdhi equations}
For the fully non--linear 1D Serre--Green Nagdhi equations we examine three test cases.  Two come from Bonneton et al. \cite{Bonneton}, and the other from Dutykh et al. \cite{Dutykh2}.  

\begin{center}
\begin{table}[h]
\centering
\renewcommand*\arraystretch{1.4}
\begin{tabular}{|l|l|l|l|}
\hline 
& One Soliton & One Soliton & One Soliton \\ 
\hline 
Amplitude $(a)$ & 0.1 & 0.025 & 0.05  \\ 
\hline 
Depth above bottom $(d)$ & 0.5 & 0.5 & 1  \\ 
\hline 
Speed $(c)$ & 2.4343 & 2.2771 & 1.0247  \\ 
\hline 
Gravity $(g)$ & 9.8765 & 9.8765 & 1  \\ 
\hline 
Final time $(T)$ & 2 & 2 & 2  \\ 
\hline 
Spatial domain length & $[-30,30]$ & $[-50,50]$ & $[-100,100]$  \\ 
\hline 
Radial basis function & Gaussian (GA) & Gaussian (GA)& Gaussian  (GA)\\ 
\hline 
RBF shape parameter & 2 & 2 & 1  \\ 
\hline 
Reference & Bonneton et al. \cite{Bonneton} & Bonneton et al. \cite{Bonneton} & Dutykh et al. \cite{Dutykh2}  \\ 
\hline 
$N$ & 400 & 400 & 400 \\ 
\hline 
Associated Figure & \ref{fig:ch_6_4_7a} and \ref{fig:ch_6_4_7b} & \ref{fig:ch_6_4_7c} and \ref{fig:ch_6_4_7d} & \ref{fig:ch_6_4_8} \\ 
\hline 
\end{tabular}
\caption[Test problems for the 1D SGN equations.]{Test problems for the 1D SGN equations.}%
\label{table_ch_6_sgn_1}
\end{table}
\end{center}

\begin{figure}[htb!]
\centering
%\hspace*{-5ex}
\begin{subfigure}{.4\textwidth}
  \centering
  \hspace*{2ex}
  \centerline{\includegraphics[trim = 10mm 80mm 20mm 85mm, clip, scale = 0.45]{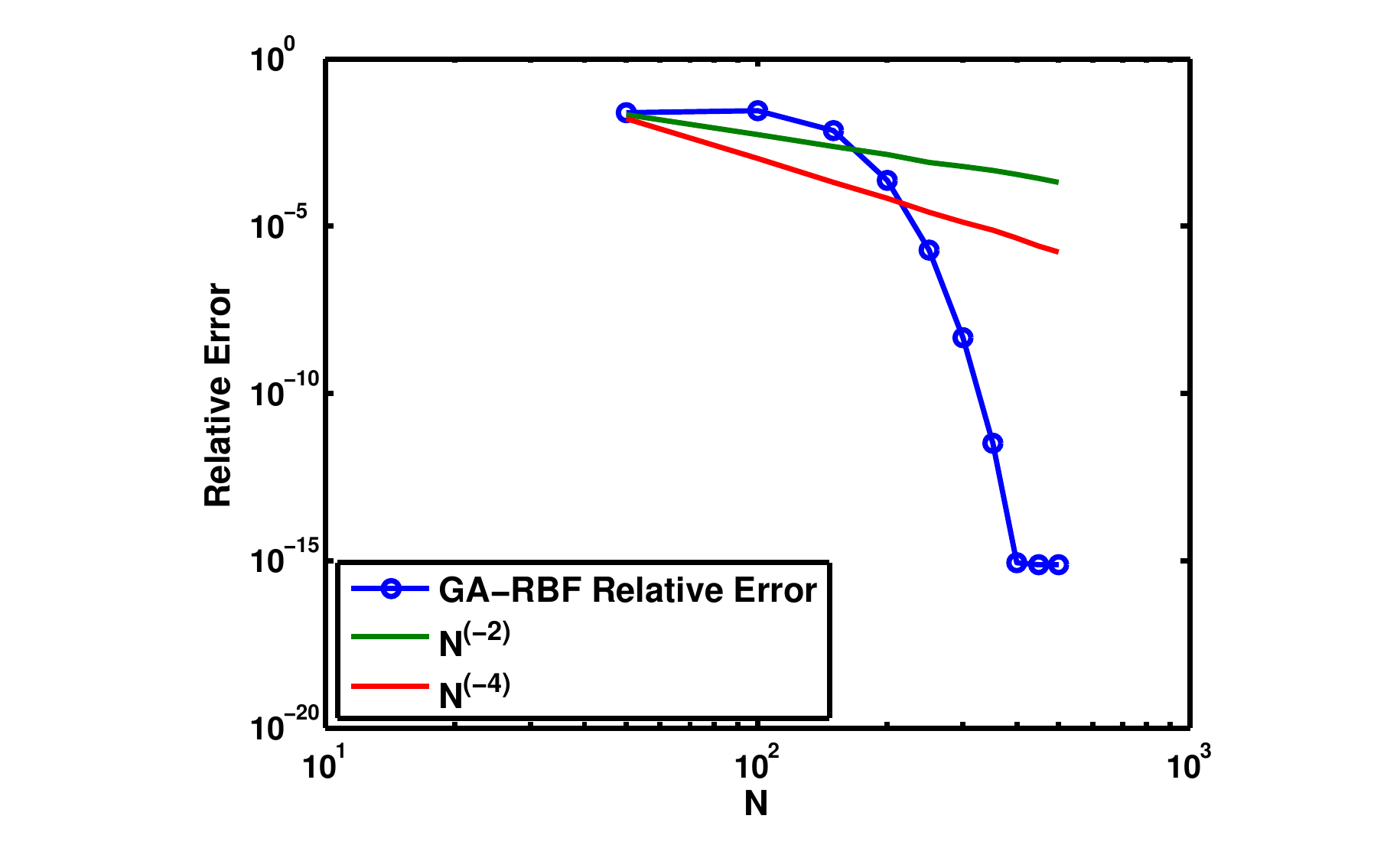}}
  \caption{ $h_0=a=0.025$, ($\eta (x,3)$)}
  \label{fig:ch_6_4_7a}
\end{subfigure}%
\hspace*{8ex}
\begin{subfigure}{.5\textwidth}
  \centering
  \centerline{\includegraphics[trim = 10mm 80mm 20mm 85mm, clip, scale = 0.45]{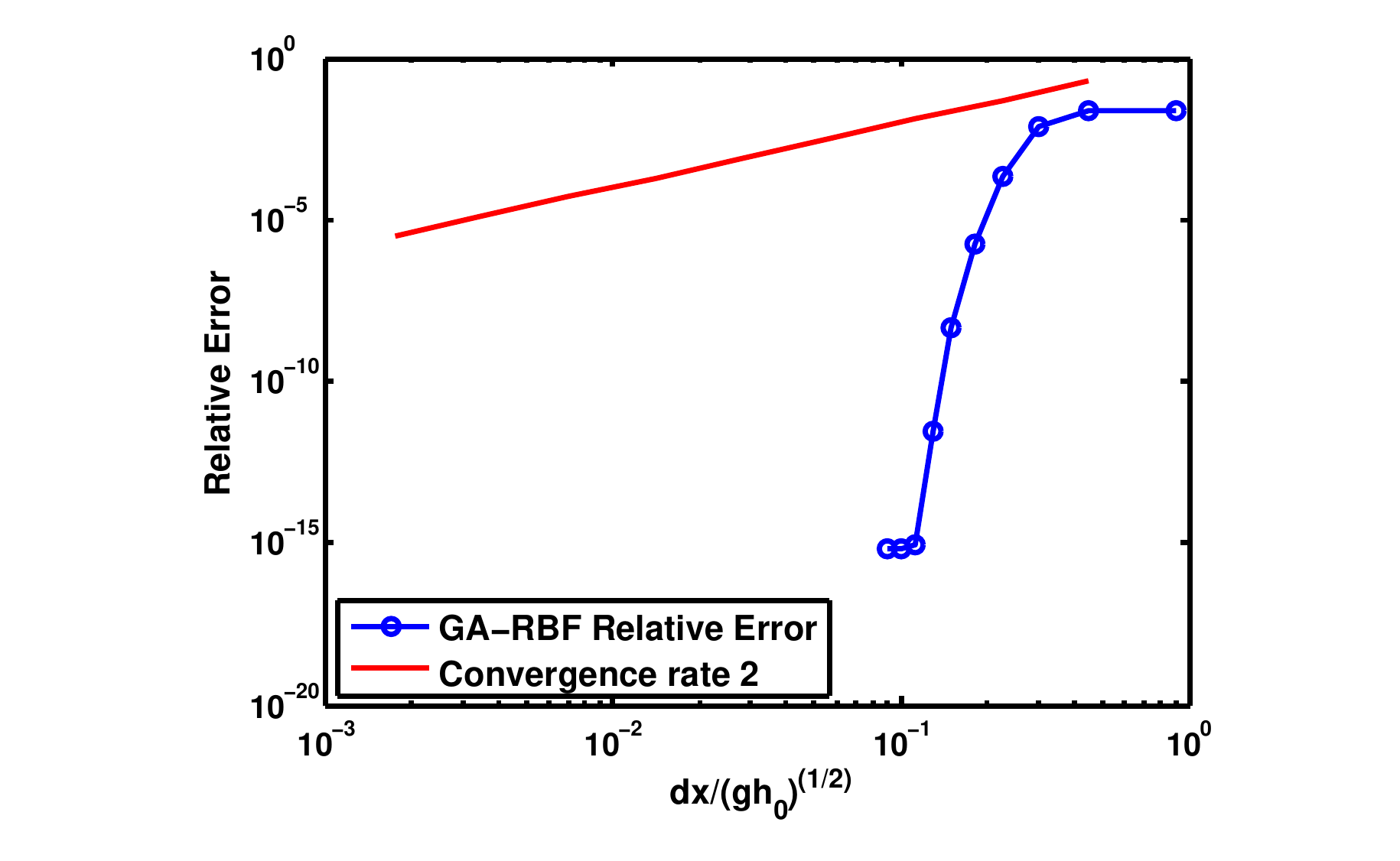}}
  \caption{ $h_0=a=0.025$, ($\eta (x,3)$)}
  \label{fig:ch_6_4_7b}
\end{subfigure}
\hspace*{-2ex}
\begin{subfigure}{.5\textwidth}
  \centering
  \centerline{\includegraphics[trim = 10mm 80mm 20mm 85mm, clip, scale = 0.45]{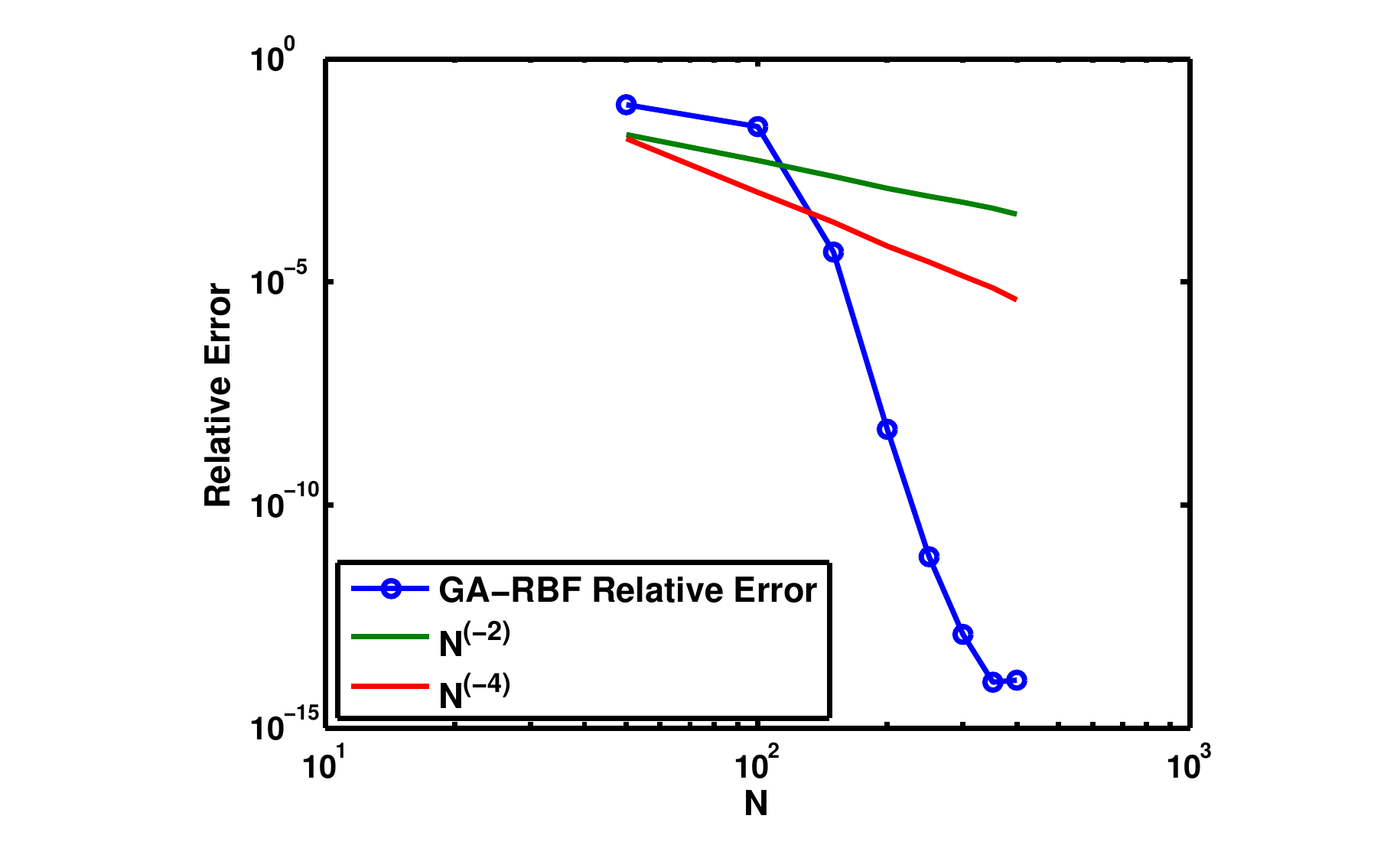}}
  \caption{ $h_0=a=0.1$, ($\eta (x,3)$)}
  \label{fig:ch_6_4_7c}
\end{subfigure}%
\hspace*{1ex}
\begin{subfigure}{.5\textwidth}
  \centering
  \centerline{\includegraphics[trim = 10mm 80mm 20mm 85mm, clip, scale = 0.45]{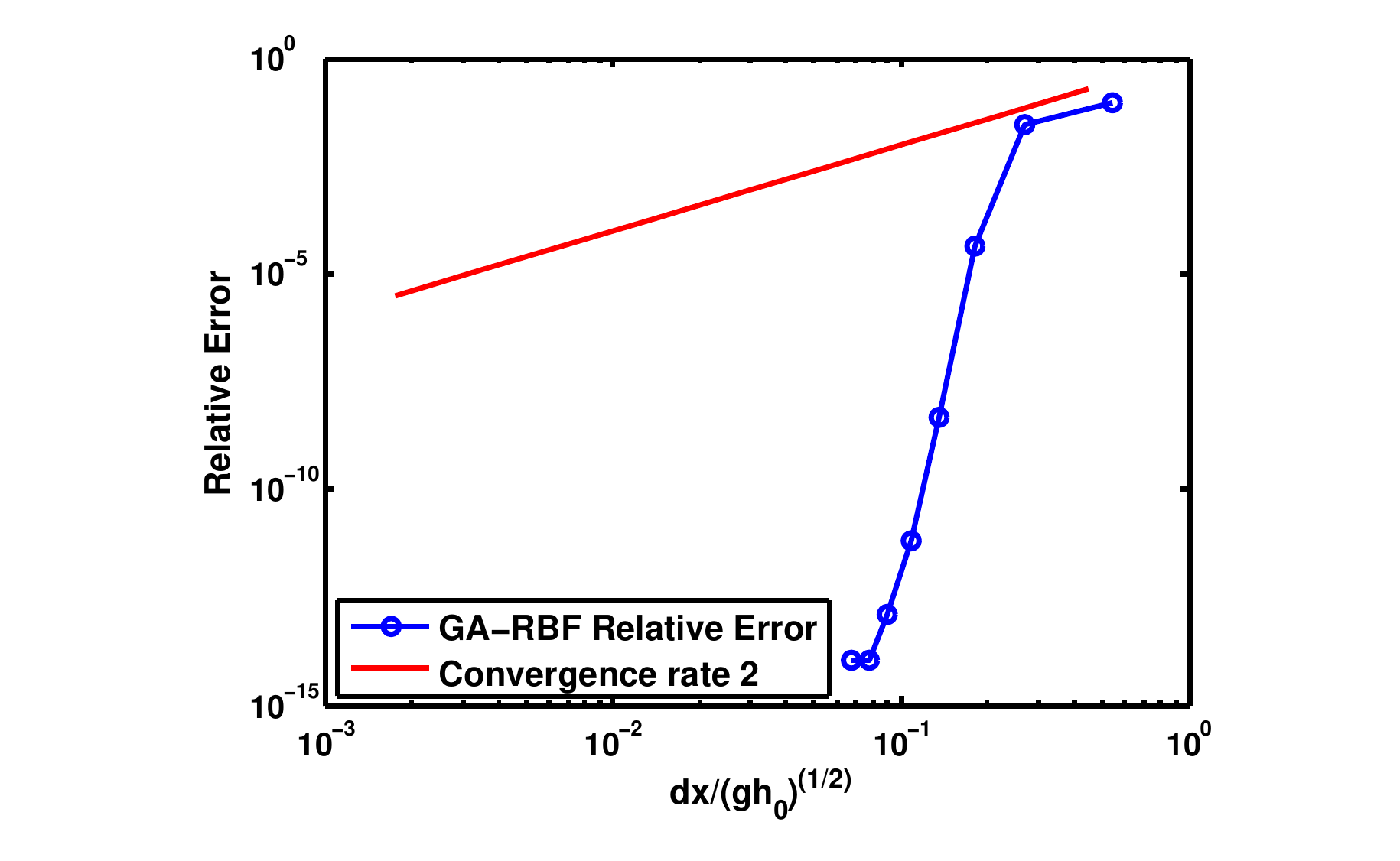}}
  \caption{ $h_0=a=0.1$ }
  \label{fig:ch_6_4_7d}
\end{subfigure}
\caption{Spatial convergence for Gaussian--RBF spectral method applied to the Bonneton et al. test cases \cite{Bonneton}.  Error is relative, $\lVert \text{approx} - \text{exact}\rVert_{\infty}/\lVert \text{exact}\rVert_{\infty}$.}
\label{fig:ch_6_4_7}
\end{figure}

\noindent  Kim in \cite{Kim} and Bonneton et al. in \cite{Bonneton} have studies of finite volume operator splitting methods applied to the 1D SGN equations.  For the test cases examined in both \cite{Kim} and \cite{Bonneton}, a second order convergence rate is observed with a Strang splitting.  Figure~\ref{fig:ch_6_4_7} confirms a near--spectral convergence rate for the same test cases described in \cite{Bonneton} (the spatial domains in table~\ref{table_ch_6_sgn_1} are slightly larger; also, since the solution decays rapidly for large $x$, zero flux boundary conditions are imposed).  The values of $N$ are much more moderate in the GA--RBF method.  For instance, the largest $\Delta x$ used in \cite{Bonneton} is $\Delta x=1/256$.  For a domain of $[-15,15]$, this corresponds to 7681 evenly spaced grid points.  The RBF spectral method uses dense, highly ill conditioned matrices; so a grid spacing of that magnitude is not tractable.  In Figure~\ref{fig:ch_6_4_7} one can see that the relative error is near machine precision for $N=500$, which corresponds to a $\Delta x = 0.0601$ on a domain of length 30.

\begin{figure}[htb!]
\begin{subfigure}{.4\textwidth}
  \centering
  \hspace*{2ex}
  \centerline{\includegraphics[trim = 10mm 80mm 20mm 85mm, clip, scale = 0.45]{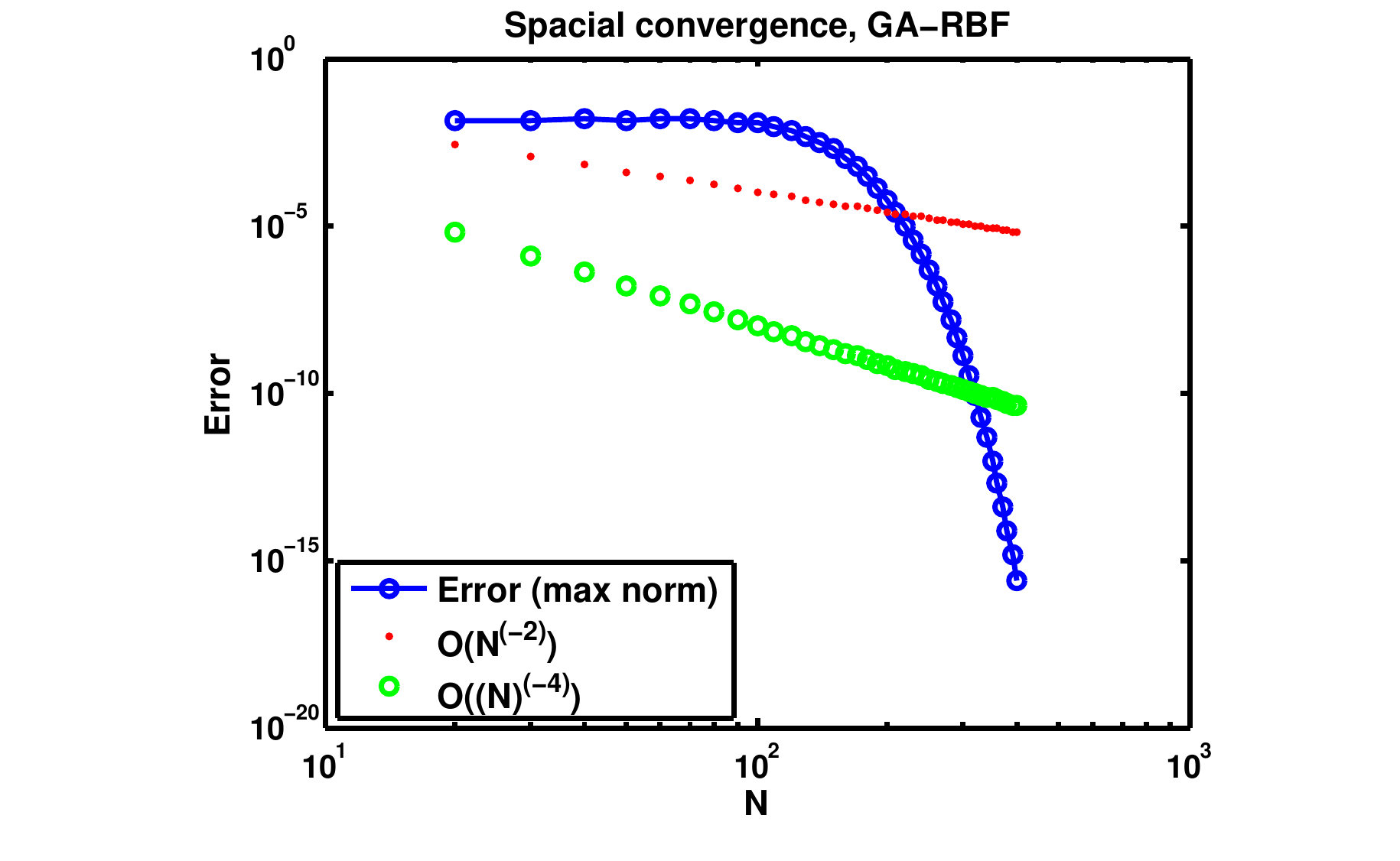}}
  \caption{ Log--log ($\eta (x,2)$) }
  \label{fig:ch_6_4_8a}
\end{subfigure}%
\hspace*{10ex}
\begin{subfigure}{.4\textwidth}
  \centering
  \centerline{\includegraphics[trim = 10mm 80mm 20mm 85mm, clip, scale = 0.45]{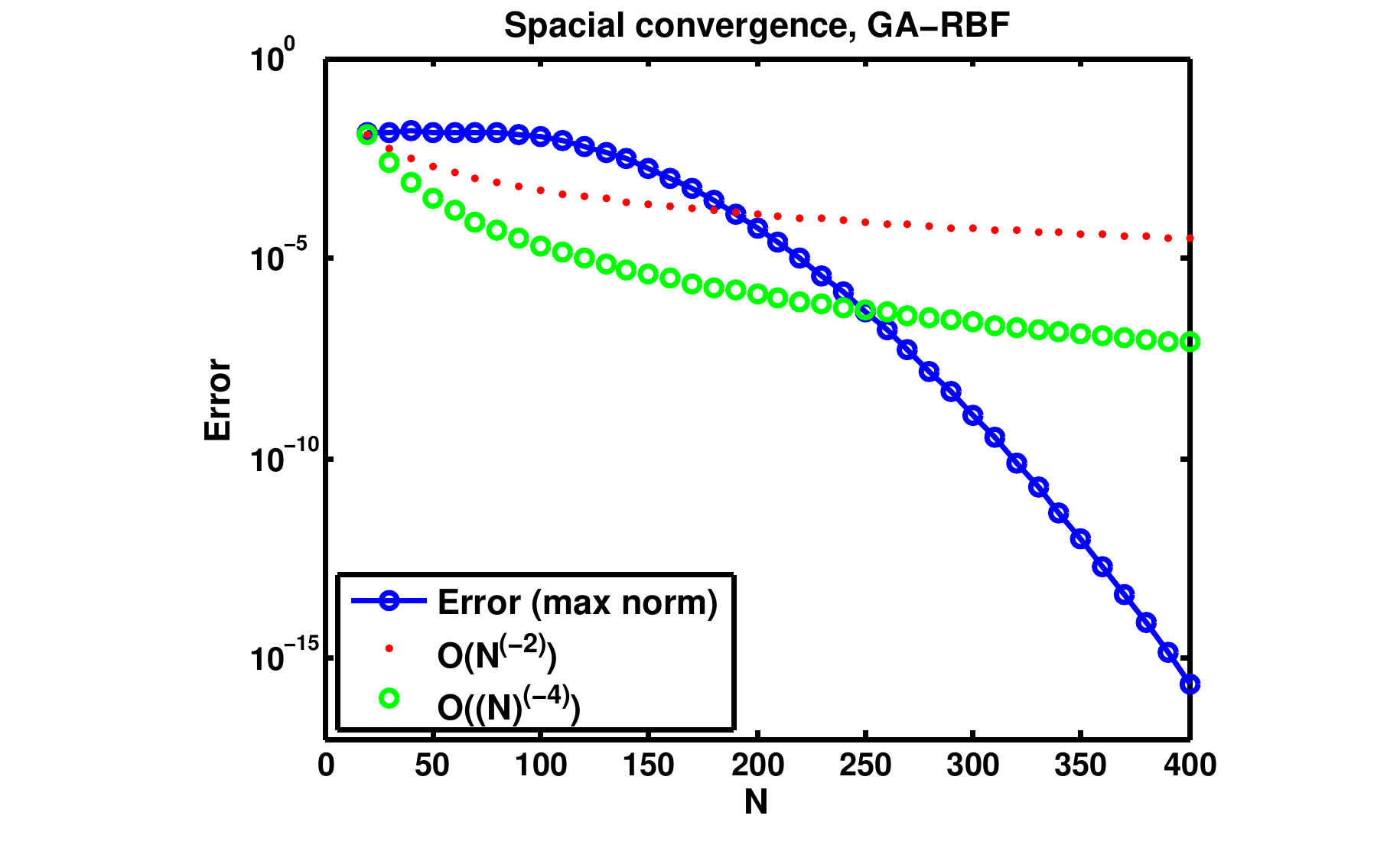}}
  \caption{ Semi--log ($\eta (x,2)$) }
  \label{fig:ch_6_4_8b}
\end{subfigure}
\caption{Spatial convergence for Gaussian--RBF spectral method applied to the Dutykh et al. test cases \cite{Dutykh2}.  Error is in the infinity norm, $\lVert \text{approx} - \text{exact}\rVert_{\infty}$.}
\label{fig:ch_6_4_8}
\end{figure}

\indent  As far as the author is aware, these results for the RBF pseudospectral method applied to the 1D SGN equations are new.  In \cite{Dutykh2} a Fourier pseudospectral method is applied to the 1D SGN equations.  Figures~\ref{fig:ch_6_4_7} and~\ref{fig:ch_6_4_8} demonstrate that the global RBF approximation is indeed a spectral method.  It is clear that spectral convergence of the spatial error occurs: as the number of grid points increases linearly, the error follows an exponential decay $\mathcal{O}(C^{-N})$ for $C>1$.

\begin{figure}[htb!]
  \centering
  \includegraphics[trim = 10mm 80mm 20mm 85mm, clip, scale = 0.55]{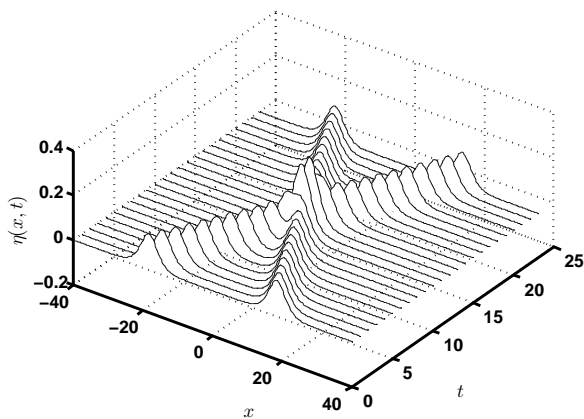}
\caption{Gaussian--RBF spectral method simulation of a head on collision.  Relevant parameters: $\epsilon=2$, $N=300$, $L=30$, $T=36$, $a=0.15$, and $x_0=\pm 20$.  This test case can also be found in \cite{Dutykh2}.}
\label{fig:ch_6_4_coll}
\end{figure}

\clearpage
\noindent  The errors measured below are for the function $\eta(x,t)$ (free surface elevation).  In Figure~\ref{fig:ch_6_4_coll}, a head on collision of two solitons is simulated.

\section{Conclusion}
We presented a RBF spectral method to the fully nonlinear one--dimensional Serre Green-Naghdi equations.  The numerical method investigated used explicit time stepping and an RBF discretization in the spatial dimesnions.  Spectral accuracy in the spatial dimension was observed for test cases in Bonneton et al. \cite{Bonneton} and Dutykh et al. \cite{Dutykh2}.  The accuracy is much higher than more robust numerical approaches based on finite volumes (which only have a second order convergence rate).  Further work includes investigating more efficient local RBFs, and possible extensions to two--dimensions.
\section*{Acknowledgments}
I would like to acknowledge useful discussions concerning this work 
with Professor Randall J. LeVeque the University of Washington.
\newpage
 \section*{Appendix A}
\begin{Verbatim}
%%%%%%%%%%%%%%%%%%%%%%%%%%%%%%%%%%%%%%%%%%%%%%%%%%%%%%%%%%%%%%%%%%%%%%%%%%%
%-% ------------------------------------------------------------------- %-%
%-% Dynamic gravity wave simulations to the one-layer                   %-%
%-% Serre-Green-Naghdi equations                                        %-%
%-% ------------------------------------------------------------------- %-%
%-% A single soliton test case is simulated. Spectral accuracy is clear %-%
%-% in the spatial dimension for n ranging linearly from n=50 to 400.   %-%
%-% (e.g. n=50:50:400)                                                  %-%
%-% ------------------------------------------------------------------- %-%
%-% Author: Maurice S. Fabien, University of Washington (Jan-Jun 2014)  %-%
%-%                          , Rice University          (2014-    )     %-%
%-% Email : fabien@rice.edu                                             %-%
%-% GitHub: https://github.com/msfabien/                                %-%
%-% ------------------------------------------------------------------- %-%
%%%%%%%%%%%%%%%%%%%%%%%%%%%%%%%%%%%%%%%%%%%%%%%%%%%%%%%%%%%%%%%%%%%%%%%%%%%
function RBF_SGN()
    clear all;  close all;  clc;  format short
    n = 200;          %For error near machiene precision, try n = 400
    Tfin = 3;         %Final time
    tspan = [0 Tfin]; %Temporal domain for ode113
    L = 50;           %Domain half length

    x  = linspace(-L,L,n)';
    Nx = length(x);
    cx = (2)*ones(Nx,1);  %blanket RBF shape parameters

    [Ax,D1x,D2x] = deal(zeros(Nx));
    for j = 1 : Nx
        [Ax(:,j),D1x(:,j),D2x(:,j)] = gau(x,x(j),cx(j));
    end
    D1x = D1x / Ax;  D2x = D2x / Ax;
    %Zero flux boundary conditions.
    D1x(1,:) = zeros(size(D1x(1,:)));  D1x(end,:) = zeros(size(D1x(1,:)));
    D2x(1,:) = zeros(size(D1x(1,:)));  D2x(end,:) = zeros(size(D1x(1,:)));

    %Various physical parameters
    d = 0.5;                  %Depth above sea floor
    g = (1/(0.45*sqrt(d)))^2; %Acceleration due to gravity
    a = 0.025;                %Soliton amplitude
    BETA = 1.0 / 3.0;
    c = sqrt(g*(d+a));        %Soliton speed
    kappa = sqrt(3*a)/(d*sqrt(a+d));

    %Exact solutions
    eta = @(x,t) a * sech( 0.5*kappa*(x - c * t)).^2;
    u   = @(x,t) c*eta(x,t) ./ (d + eta(x,t));
    q   = @(x,t) u(x,t) - (d+eta(x,t)).*((BETA)*(d+eta(x,t)).*(D2x*(u(x,t)))...
        + (D1x*(eta(x,t))).*(D1x*(u(x,t))));

    %Initial conditions
    Q = q(x,0);  ETA = eta(x,0.0);
    init = [ETA; Q];
    %strict ode suite error tolerances
    options = odeset('RelTol',2.3e-14,'AbsTol',eps);
    tic
    [t,w] = ode113(@(t,q) RHS(t,q,0,D1x,D2x,g,d,BETA), tspan,init,options);
    toc

    %Error analysis
    ETA = w(end,1:Nx)';   %RBF spectral approximation for \eta(x,3)
    Q = w(end,Nx+1:end)'; %RBF spectral approximation for    q(x,3)
    L1 = BETA*diag((d+ETA).^2)*D2x+diag((d+ETA).*(D1x*eta(x,Tfin)))*D1x-eye(n);
    U = L1 \ (-Q);    %RBF spectral approximation for    u(x,3)

    Exact_Error_ETA    = norm( ETA - eta(x,Tfin) , inf) 
    Relative_Error_ETA = norm( ETA - eta(x,Tfin) , inf)/norm( eta(x,Tfin) , inf)

    Exact_Error_U    = norm( U - u(x,Tfin) , inf) 
    Relative_Error_U = norm( U - u(x,Tfin) , inf)/norm( u(x,Tfin) , inf)

    Exact_Error_Q    = norm( Q - q(x,Tfin) , inf) 
    Relative_Error_Q = norm( Q - q(x,Tfin) , inf)/norm( q(x,Tfin) , inf)
    plot(x,eta(x,0.0),'g',x,eta(x,Tfin),'b',x,ETA,'r.')
    xlabel('x'),  ylabel('\eta')
    legend('\eta(x,0)','\eta(x,3)','RBF approximation','Location','best')
end

function [phi,phi1,phi2] = gau(x,xc,c)
    % Computes 1-D guassian radial basis function interpolation and
    % differentation matrices.  Higher order derivatives need to be added.
    f = @(r,c) exp(-(c*r).^2);
    r = x - xc;
    phi = f(r,c);
    if nargout > 1  
	% 1st derivative  
        phi1 = -2*r*c^2.*exp(-(c*r).^2);  
        if nargout > 2
        % 2nd derivative
            phi2 = 2*c^2*exp(-c^2*r.^2).*(2*c^2*r.^2 - 1);
        end
    end
end

function f = RHS(t,q,dummy,D1x,D2x,g,d,BETA) 
    % Right hand side function for ODE solver
    n  = length(D1x);  I = eye(n);  
    ETA  = q(1:n);     ETA_x = D1x*ETA;  Q = q(n+1 : 2*n);

    % Linear system solved for U (comes from elliptic equation)
    L   = ( BETA*diag((d+ETA).^2)*D2x + diag((d+ETA).*ETA_x)*D1x - I );
    U   = L \ (-Q);

    rhs1 = -D1x*((d+ETA) .*U );
    rhs2 = -D1x*(Q.*U - 0.5*(U).^2 + g*ETA - 0.5*(d+ETA).^2.*((D1x*U).^2));
    f = [rhs1; rhs2];
end
\end{Verbatim}
 
\bibliographystyle{spmpsci}
\bibliography{biblo}

\end{document}